\documentclass[english]{IEEEtran}
\usepackage[T1]{fontenc}
\usepackage[latin9]{inputenc}
\usepackage{xcolor}
\usepackage{pdfcolmk}
\usepackage{float}
\usepackage{calc}
\usepackage{bm}
\usepackage{amsmath}
\usepackage{amsthm}
\usepackage{amssymb}
\usepackage{graphicx}
\usepackage{setspace}
\PassOptionsToPackage{normalem}{ulem}
\usepackage{ulem}

\makeatletter

\providecommand{\tabularnewline}{\\}
\floatstyle{ruled}
\newfloat{algorithm}{tbp}{loa}
\providecommand{\algorithmname}{Algorithm}
\floatname{algorithm}{\protect\algorithmname}
\providecolor{lyxadded}{rgb}{0,0,1}
\providecolor{lyxdeleted}{rgb}{1,0,0}
\DeclareRobustCommand{\lyxsout}[1]{\ifx\\#1\else\sout{#1}\fi}

\theoremstyle{definition}
\newtheorem{defn}{\protect\definitionname}
\theoremstyle{plain}
\newtheorem{thm}{\protect\theoremname}
\theoremstyle{plain}
\newtheorem{lem}{\protect\lemmaname}
\theoremstyle{remark}
\newtheorem{rem}{\protect\remarkname}

\usepackage{cite}
\usepackage{times}
\newtheorem{assumption}{Assumption}
\newtheorem{challenge}{Challenge}

\makeatother

\usepackage{babel}
\providecommand{\definitionname}{Definition}
\providecommand{\lemmaname}{Lemma}
\providecommand{\remarkname}{Remark}
\providecommand{\theoremname}{Theorem}

\begin{document}

\title{Two-Timescale Hybrid Compression and Forward for Massive MIMO Aided
C-RAN}

\author{{\normalsize{}An Liu, }\textit{\normalsize{}Senior Member, IEEE}{\normalsize{},
Xihan Chen, Wei Yu, }\textit{\normalsize{}Fellow, IEEE}{\normalsize{},
Vincent Lau,}\textit{\normalsize{} Fellow, IEEE}{\normalsize{} and
Min-Jian Zhao, }\textit{\normalsize{}Member, IEEE}{\normalsize{}}\thanks{This work was supported by the Science and Technology Program of Shenzhen,
China, under Grant JCYJ20170818113908577, the National Natural Science
Foundation of China under Project No. 61571383, and RGC 16209916.
The work of An Liu was supported by the China Recruitment Program
of Global Young Experts.

An Liu, Xihan Chen and Min-Jian Zhao are with the College of Information
Science and Electronic Engineering, Zhejiang University, Hangzhou
310027, China (e-mail: anliu@zju.edu.cn, chenxihan@zju.edu.cn, mjzhao@zju.edu.cn).

Wei Yu is with the Electrical and Computer Engineering Department,
University of Toronto, Toronto, ON M5S 3G4, Canada (e-mail: weiyu@ece.utoronto.ca).

Vincent Lau is with the Department of ECE, The Hong Kong University
of Science and Technology (email: eeknlau@ece.ust.hk).}}
\maketitle
\begin{abstract}
We consider the uplink of a cloud radio access network (C-RAN), where
massive MIMO remote radio heads (RRHs) serve as relays between users
and a centralized baseband unit (BBU). Although employing massive
MIMO at RRHs can improve the spectral efficiency, it also significantly
increases the amount of data transported over the fronthaul links
between RRHs and BBU, which becomes a performance bottleneck. Existing
fronthaul compression methods for conventional C-RAN are not suitable
for the massive MIMO regime because they require fully-digital processing
and/or real-time full channel state information (CSI), incurring high
implementation cost for massive MIMO RRHs. To overcome this challenge,
we propose to perform a two-timescale hybrid analog-and-digital spatial
filtering at each RRH to reduce the fronthaul consumption. Specifically,
the analog filter is adaptive to the channel statistics to achieve
massive MIMO array gain, and the digital filter is adaptive to the
instantaneous effective CSI to achieve spatial multiplexing gain.
Such a design can alleviate the performance bottleneck of limited
fronthaul with reduced hardware cost and power consumption, and is
more robust to the CSI delay. We propose an online algorithm for the
two-timescale non-convex optimization of analog and digital filters,
and establish its convergence to stationary solutions. Finally, simulations
verify the advantages of the proposed scheme.
\end{abstract}

\begin{IEEEkeywords}
C-RAN, Massive MIMO, Hybrid compression and forward, Two-stage stochastic
optimization

\thispagestyle{empty}
\end{IEEEkeywords}

\section{Introduction}

Cloud radio access network (C-RAN) \cite{CMRI_TR11_CRAN} and massive
multiple-input multiple-output (MIMO) \cite{Marzetta_SPM12_LargeMIMO}
are regarded as two key technologies for future wireless systems.
Both technologies can significantly improve the spectral and energy
efficiency of wireless systems by employing a huge number of antennas
per unit area. However, they adopt different architectures and thus
have their own pros and cons. 

C-RAN is essentially a large-scale distributed antenna system, where
plenty of remote radio heads (RRHs) are distributed within a specific
geographical area and are connected to a centralized baseband unit
(BBU) pool through fronthaul links. Each RRH merely serves as a relay
to forward the signals from/to the BBU via its fronthaul link, while
all baseband processings are performed at the BBU. Since each user
can always find some nearby RRHs with strong channel conditions, the
users at different locations can enjoy a uniform quality of experience
without suffering from the cell-edge effect. However, in practice,
the performance of C-RAN is limited by the fronthaul capacity between
each RRH's and BBU, especially when each RRH has multiple antennas.
In contrast, the massive MIMO system deploys a large number of antennas
at the base station (BS) to achieve large spatial multiplexing and
array gains. In this case, processing is done locally at the BS, hence
the performance is no longer limited by the fronthaul capacity.

Recently, massive MIMO aided C-RAN, in which each RRH is equipped
with a massive MIMO array, has been proposed to further improve the
spectral and energy efficiency of wireless systems \cite{Chen_WCM17_HCRAN}.
However, moving signal processing of an uplink massive MIMO system
from the RRH to the cloud would require a huge amount of digital sampled
data to be transported over the fronthaul link. Therefore, it is necessary
to compress the uplink data at each RRH to satisfy the limited fronthaul
capacity constraint. Various fully-digital fronthaul compression techniques
have been proposed for the uplink of C-RAN with small-scale multi-antenna
RRHs, from the more complicated quantize-and-forward (QF) schemes
\cite{Park_VTC2013_QFCRAN,Zhou_JSAC14_QFCRAN} to the simpler uniform
scalar quantization schemes \cite{Liu_TCOM15_UQCRAN} and RRH selection
schemes \cite{Luo_TWC15_RselCRAN}. In particular, the spatial compression
and forward scheme proposed in \cite{Liu_TSP2015_SCFCRAN} combines
fully-digital linear spatial filtering and uniform scalar quantization
to alleviate the performance bottleneck caused by the limited fronthaul
capacity. Unfortunately, fully-digital spatial filtering requires
a larger number of analog-to-digital converter (ADCs) and radio frequency
(RF) chains at each massive MIMO RRH. In \cite{Combi_EuCNC2017_FACRAN},
a fully-analog linear spatial filtering is used at each RRH to achieve
the fronthaul compression with reduced hardware cost and power consumption.
However, fully-analog processing is known to be less efficient than
hybrid analog and digital processing. Moreover, the analog filtering
matrix in \cite{Combi_EuCNC2017_FACRAN} is adapted to the instantaneous
channel state information (CSI), making it difficult to be extended
to wideband systems with many subcarriers, because the instantaneous
CSI on different subcarriers is usually different \cite{Liu_TSP2016_CSImassive}.

In this paper, we propose a two-timescale hybrid (analog and digital)
compression and forward (THCF) scheme for the uplink transmission
of massive MIMO aided C-RAN, to alleviate the performance bottleneck
of the limited fronthaul, with reduced hardware cost and power consumption.
In this scheme, each RRH first performs a two-timescale hybrid analog
and digital spatial filtering to reduce the dimension of its received
signal. Specifically, the analog filtering matrix is adapted to the
long-term channel statistics to achieve massive MIMO array gain, and
the digital filtering matrix is adapted to the instantaneous effective
CSI (i.e., the product of the instantaneous channel and analog filtering
matrix) to achieve spatial multiplexing gain. Then, each RRH applies
the uniform scalar quantization over each of these dimensions. Finally,
the quantized signals at the RRHs are sent to the BBU for joint decoding.
The power allocation at users, analog/digital filtering matrices and
quantization bits allocation at RRHs, as well as the receive beamforming
matrix at the BBU are jointly optimized to maximize a general utility
function of long-term average data rates of users, including average
weighted sum-rate maximization and proportional fairness (PFS) utility
maximization as special cases. 

Such a two-timescale hybrid design has several advantages. For example,
the analog filtering matrix is robust to the CSI signaling latency.
Moreover, since the channel statistics is approximately the same over
different subcarriers \cite{Sadek_TOC08_chstatistic}, a single analog
filtering matrix is sufficient to cover all subcarriers at each RRH,
making it applicable to wideband systems. With the proposed THCF scheme,
the massive MIMO aided C-RAN uplink system can enjoy the huge array
gain provided by the massive MIMO almost for free (i.e., the complexity
and power consumption are similar to the C-RAN with small-scale multi-antenna
RRHs). However, there are also several technical challenges in the
implementation of this architecture.
\begin{itemize}
\item \textbf{Two-timescale Stochastic Non-convex Optimization:} The joint
optimization of long-term control variables (analog filtering) and
short-term control variables (power allocation, digital filtering,
quantization bits allocation, and receive beamforming matrix) belongs
to two-timescale stochastic non-convex optimization, which is difficult
to solve. Specifically, the objective function contains expectation
operators and the argument of the expectation operators involves the
optimal short-term control variables, which do not have closed-form
expressions. In addition, the optimization of the short-term control
variables at different time slots are usually coupled together for
a general utility function such as PFS. Moreover, both short-term
and long-term subproblems are non-convex.
\item \textbf{Lack of Channel Statistics:} In practice, we may not even
have explicit knowledge of the channel statistics. Hence, the solution
should be self-learning to the unknown channel statistics.
\item \textbf{Convergence Analysis: }It is very important to establish the
convergence of the algorithm. However, this is non-trivial for a two-timescale
stochastic non-convex optimization problem. 
\end{itemize}

To address the above challenges, we propose an online \textit{block-coordinate
stochastic successive convex approximation} (BC-SSCA) algorithm with
self-learning capability to solve the two-timescale stochastic non-convex
optimization problem without explicit knowledge of the channel statistics.
In addition, we establish convergence of the BC-SSCA algorithm to
stationary solutions. Finally, simulations show that the proposed
two-timescale hybrid scheme achieves better tradeoff performance than
the baselines. 

The rest of the paper is organized as follows. In Section \ref{sec:System-Model},
we give the system model for two-timescale hybrid compression and
forward in the uplink of massive MIMO aided C-RAN. In Section \ref{sec:Two-timescale-Joint-Optimization},
we formulate the two-timescale stochastic non-convex optimization
problem for the joint optimization of long-term and short-term control
variables. The proposed online BC-SSCA algorithm and the associated
convergence proof are presented in Section \ref{sec:Constrained-Stochastic-Successiv}.
The simulation results are given in Section \ref{sec:Simulation-Results}
to verify the advantages of the proposed solution, and the conclusion
is given in Section \ref{sec:Conlusion}. The key notations used in
this paper are summarized in Table \ref{notation}.

\begin{table}
\begin{centering}
\begin{tabular}{|c|l|}
\hline 
Symbol & Parameters\tabularnewline
\hline 
\hline 
{\small{}$N$ ($n$)} & {\small{}Number of RRHs (index for RRH)}\tabularnewline
\hline 
{\small{}$M$} & {\small{}Number of antennas at each RRH}\tabularnewline
\hline 
{\small{}$S$} & {\small{}Number of RF chains at each RRH}\tabularnewline
\hline 
{\small{}$K$ ($k$)} & {\small{}Number of users (index for user)}\tabularnewline
\hline 
{\small{}$L$} & {\small{}Signal dimension after compression }\tabularnewline
\hline 
{\small{}$l$} & {\small{}Index for the entry of compressed signal}\tabularnewline
\hline 
{\small{}$i$} & {\small{}Index for time slot}\tabularnewline
\hline 
{\small{}$t$} & {\small{}Index for frame}\tabularnewline
\hline 
{\small{}$\bm{v}$} & {\small{}Digital filtering vector (short-term)}\tabularnewline
\hline 
{\small{}$\bm{d}$} & {\small{}Quantization bits allocation (short-term)}\tabularnewline
\hline 
{\small{}$\bm{u}$} & {\small{}Rx beamforming vector (short-term)}\tabularnewline
\hline 
{\small{}$\bm{p}$} & {\small{}Transmit power vector (short-term)}\tabularnewline
\hline 
{\small{}$\bm{\theta}$} & {\small{}Phase vector of analog filtering (long-term)}\tabularnewline
\hline 
{\small{}$\Theta$} & {\small{}Feasible set of $\bm{\theta}$}\tabularnewline
\hline 
{\small{}$\mathcal{X}$ } & {\small{}Feasible set of the short-term variables}\tabularnewline
\hline 
{\small{}$\widetilde{\mathcal{X}}$} & {\small{}Relaxed feasible set of the short-term variables}\tabularnewline
\hline 
{\small{}$\Omega$} & {\small{}Collection of short-term variables}\tabularnewline
\hline 
{\small{}($r_{k}$) $r_{k}^{\circ}$ } & {\small{}(Approximate) data rate of user $k$}\tabularnewline
\hline 
{\small{}($\overline{r}_{k}$) $\overline{r}_{k}^{\circ}$} & {\small{}(Approximate) average data rate of user $k$}\tabularnewline
\hline 
{\small{}$g\left(\overline{\boldsymbol{r}}\right)$} & {\small{}Utility function}\tabularnewline
\hline 
{\small{}$\hat{r}_{k}^{t}$} & {\small{}Recursive approximation for $\overline{r}_{k}$}\tabularnewline
\hline 
{\small{}$\mathbf{f}^{t}$} & {\small{}Recursive approximation for $\nabla_{\boldsymbol{\theta}}g\left(\overline{\boldsymbol{r}}\right)$}\tabularnewline
\hline 
\end{tabular}
\par\end{centering}
\caption{\label{notation}List of notations.}
\end{table}

\section{System Model\label{sec:System-Model}}

\subsection{Network Architecture and Channel Model}

Consider the uplink of a massive MIMO aided C-RAN, where $N$ RRHs,
each equipped with a massive MIMO array of $M\gg1$ antennas and $S<M$
Rx RF chains, are distributed within a specific geographical area
to serve $K$ single-antenna users, as shown in Fig. \ref{sec:System-Model}.
Each RRH $n$ serves as a relay between the BBU and users, and is
connected to the BBU via a fronthaul link of capacity $C_{n}$ bits
per second (bps). The BBU is in charge of making resource allocation
decisions and joint decoding of the users\textquoteright{} messages
based on the signals from all the RRHs. We assume that the number
of users $K$ is fixed and $NS\gg K$ so that there are enough spatial
degrees of freedom to serve all the $K$ users. This is a typical
operating regime that has been assumed in many works on massive MIMO
systems \cite{Ayach_TWC14_mmRFprecoding,Liu_TSP2016_CSImassive,Park_TSP17_THP}.
As a motivating example, consider a system in which the users are
a fixed number of pico BSs and the RRH provides backhaul links between
the pico-cells and BBU.

\begin{figure}
\begin{centering}
\includegraphics[width=80mm]{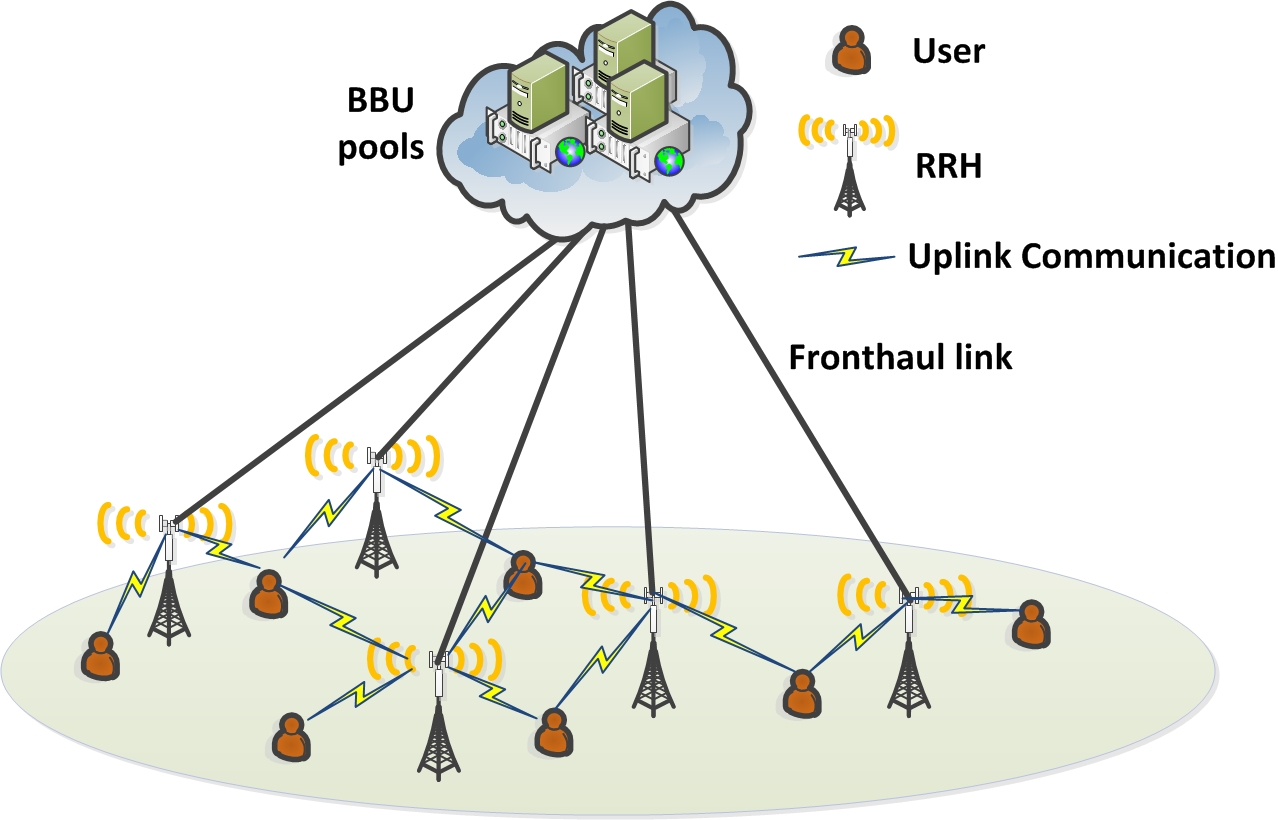}
\par\end{centering}
\caption{\label{fig:systemmodel}Uplink of a Massive MIMO aided C-RAN}
\end{figure}

For clarity, we focus on a narrowband system with flat block fading
channel, but the proposed algorithm can be easily modified to cover
the wideband system as well. In this case, the received signal at
RRH $n$ is given by
\[
\boldsymbol{y}_{n}=\sum_{k=1}^{K}\boldsymbol{h}_{n,k}\sqrt{p_{k}}s_{k}+\boldsymbol{z}_{n}=\boldsymbol{H}_{n}\boldsymbol{P}^{1/2}\boldsymbol{s}+\boldsymbol{z}_{n},
\]
where $\boldsymbol{H}_{n}=\left[\boldsymbol{h}_{n,1},...,\boldsymbol{h}_{n,K}\right]\in\mathbb{C}^{M\times K}$
with $\boldsymbol{h}_{n,k}\in\mathbb{C}^{M}$ denoting the channel
vector from user $k$ to RRH $n$, $\boldsymbol{s}=\left[s_{1},...,s_{K}\right]^{T}$
with $s_{k}\sim\mathcal{CN}\left(0,1\right)$ denoting the data symbol
of user $k$, $\boldsymbol{P}=\text{diag}\left(p_{1},...,p_{K}\right)$
with $p_{k}$ denoting the transmit power of user $k$, and $\boldsymbol{z}_{n}\sim\mathcal{CN}\left(0,\boldsymbol{I}\right)$
is the additive white Gaussian noise vector.

\subsection{Two-timescale Hybrid Compression and Forward at RRHs}

\begin{figure}
\begin{centering}
\includegraphics[viewport=0bp 0bp 400bp 170bp,width=0.95\columnwidth]{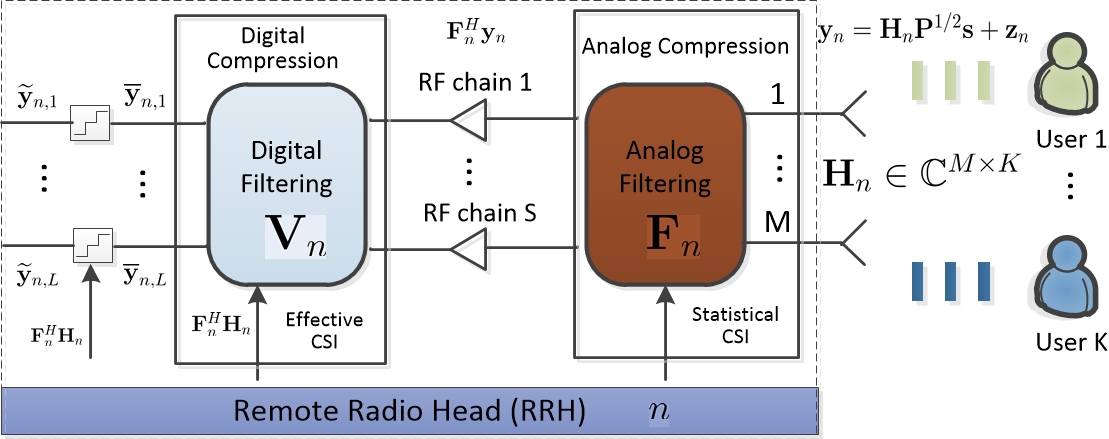}
\par\end{centering}
\caption{\label{fig:THCF}An illustration of the THCF scheme in Massive MIMO
aided C-RAN}
\end{figure}

Each RRH $n$ applies the THCF scheme to make sure that the compressed
received signal $\widetilde{\boldsymbol{y}}_{n}$ can be forward to
the BBU via its fronthaul with a limited capacity of $C_{n}$ bps,
as illustrated in Fig. \ref{fig:THCF}. Specifically, a two-timescale
hybrid filtering matrix $\boldsymbol{F}_{n}\boldsymbol{V}_{n}\in\mathbb{C}^{M\times L}$
is first applied at RRH $n$ to compress the received signal $\boldsymbol{y}_{n}$
into a low-dimensional signal $\overline{\boldsymbol{y}}_{n}=\boldsymbol{V}_{n}^{H}\boldsymbol{F}_{n}^{H}\boldsymbol{y}_{n}=\left[\overline{y}_{n,1},...,\overline{y}_{n,L}\right]^{T}\in\mathbb{C}^{L}$,
where $\boldsymbol{F}_{n}\in\mathbb{C}^{M\times S}$ and $\boldsymbol{V}_{n}=\left[\boldsymbol{v}_{n,1},...,\boldsymbol{v}_{n,L}\right]\in\mathbb{C}^{S\times L}$
are the analog and digital filtering matrices, respectively, and we
set $L=\mathrm{min}(K,S)$ such that there is no information loss
due to digital filtering at each RRH \cite{Liu_TSP2015_SCFCRAN}.
The analog filtering matrix $\boldsymbol{F}_{n}$ is usually implemented
using an RF phase shifting network \cite{Zhang_TSP05_RFshifter}.
Hence, $\boldsymbol{F}_{n}$ can be represented by a phase vector
$\boldsymbol{\theta}_{n}\in\left[0,2\pi\right]^{MS}$, whose $\left(\left(j-1\right)M+i\right)$-th
element $\theta_{n,i,j}$ is the phase of the $\left(i,j\right)$-th
element of $\boldsymbol{F}_{n}$, i.e., $\left[\boldsymbol{F}_{n}\right]_{i,j}=\frac{1}{\sqrt{M}}e^{\sqrt{-1}\theta_{n,i,j}}$.
In this paper, we assume that high-resolution ADCs are used at each
RRH such that the quantization error due to ADCs is negligible. Then,
a simple uniform scalar quantization \cite{Liu_TCOM15_UQCRAN} is
applied to each element of $\overline{\boldsymbol{y}}_{n}$ at RRH
$n$ to achieve fronthaul compression. Note that the quantization
is performed at the baseband after the digital filter instead of at
the ADC because we need to dynamically adjust the quantization bits
according to the instantaneous channel state to improve the efficiency
of fronthaul compression.

After the uniform scalar quantization, the compressed received signal
$\widetilde{\boldsymbol{y}}_{n}=\left[\widetilde{y}_{n,1},...,\widetilde{y}_{n,L}\right]^{T}$
is modeled by
\[
\widetilde{\boldsymbol{y}}_{n}=\overline{\boldsymbol{y}}_{n}+\boldsymbol{e}_{n}=\boldsymbol{V}_{n}^{H}\boldsymbol{F}_{n}^{H}\left(\boldsymbol{H}_{n}\boldsymbol{P}^{1/2}\boldsymbol{s}+\boldsymbol{z}_{n}\right)+\boldsymbol{e}_{n},
\]
where $\boldsymbol{e}_{n}=\left[e_{n,1},...,e_{n,L}\right]\in\mathbb{C}^{L}$
with $e_{n,l}$ denoting the quantization error for $\overline{y}_{n,l}$.
Let $d_{n,l}$ denote the number of bits that RRH $n$ uses to quantize
the real or imaginary part of $\overline{y}_{n,l}$. With uniform
scalar quantization, the covariance matrix of $\boldsymbol{e}_{n}$
is given by a function of $\boldsymbol{p}=\left[p_{1},...,p_{K}\right]^{T}$,
$\boldsymbol{F}_{n}\boldsymbol{V}_{n}$ and $\boldsymbol{d}_{n}=\left[d_{n,1},...,d_{n,L}\right]^{T}$
as \cite{Liu_TCOM15_UQCRAN}
\[
\boldsymbol{Q}_{n}\left(\boldsymbol{p},\boldsymbol{F}_{n}\boldsymbol{V}_{n},\boldsymbol{d}_{n}\right)=\text{diag}\left(q_{n,1},...,q_{n,L}\right),
\]
where $q_{n,l}$ is the variance of the quantization error $e_{n,l}$:
\begin{equation}
q_{n,l}=\begin{cases}
\frac{3}{4^{d_{n,l}}}(\sum_{k=1}^{K}p_{k}|\boldsymbol{h}_{n,k}^{H}\widetilde{\boldsymbol{v}}_{n,l}|^{2}+\left\Vert \widetilde{\boldsymbol{v}}_{n,l}\right\Vert ^{2}) & \text{if }d_{n,l}>0,\\
\infty & \text{if }d_{n,l}=0,
\end{cases}\label{eq:qeq}
\end{equation}
where $\widetilde{\boldsymbol{v}}_{n,l}=\boldsymbol{F}_{n}\boldsymbol{v}_{n,l}$.
Finally, each RRH forwards the quantized bits to the BBU via the fronthaul
link.

\subsection{Joint Rx Beamforming at the BBU}

The received signal $\widetilde{\boldsymbol{y}}=\left[\widetilde{\boldsymbol{y}}_{1}^{T},...,\widetilde{\boldsymbol{y}}_{N}^{T}\right]^{T}$
at the BBU from all RRHs can be expressed as
\[
\widetilde{\boldsymbol{y}}=\widetilde{\boldsymbol{V}}^{H}\boldsymbol{H}\boldsymbol{P}^{1/2}\boldsymbol{s}+\widetilde{\boldsymbol{V}}^{H}\boldsymbol{z}+\boldsymbol{e},
\]
where $\widetilde{\boldsymbol{V}}=\text{diag}\left(\boldsymbol{F}_{1}\boldsymbol{V}_{1},...,\boldsymbol{F}_{N}\boldsymbol{V}_{N}\right)\in\mathbb{C}^{MN\times LN}$,
$\boldsymbol{H}=\left[\boldsymbol{h}_{1},...,\boldsymbol{h}_{K}\right]\in\mathbb{C}^{MN\times K}$
with $\boldsymbol{h}_{k}=\left[\boldsymbol{h}_{1,k}^{T},...,\boldsymbol{h}_{N,k}^{T}\right]^{T}$
denoting the composite channel vector of user $k$, $\boldsymbol{z}=\left[\boldsymbol{z}_{1}^{T},...,\boldsymbol{z}_{N}^{T}\right]^{T}$,
and $\boldsymbol{e}=\left[\boldsymbol{e}_{1}^{T},...,\boldsymbol{e}_{N}^{T}\right]^{T}$.
A joint Rx beamforming vector $\boldsymbol{u}_{k}\in\mathbb{C}^{NL\times1}$
is applied at the BBU to obtain the estimated data symbol for each
user $k$ as 
\begin{align*}
\hat{s}_{k} & =\boldsymbol{u}_{k}^{H}\widetilde{\boldsymbol{y}}\\
 & =\boldsymbol{u}_{k}^{H}\widetilde{\boldsymbol{V}}^{H}\boldsymbol{H}\boldsymbol{P}^{1/2}\boldsymbol{s}+\boldsymbol{u}_{k}^{H}\widetilde{\boldsymbol{V}}^{H}\boldsymbol{z}+\boldsymbol{u}_{k}^{H}\boldsymbol{e},\forall k.
\end{align*}

\subsection{Frame Structure and Achievable Data Rate}

\begin{figure}
\begin{centering}
\includegraphics[width=0.95\linewidth]{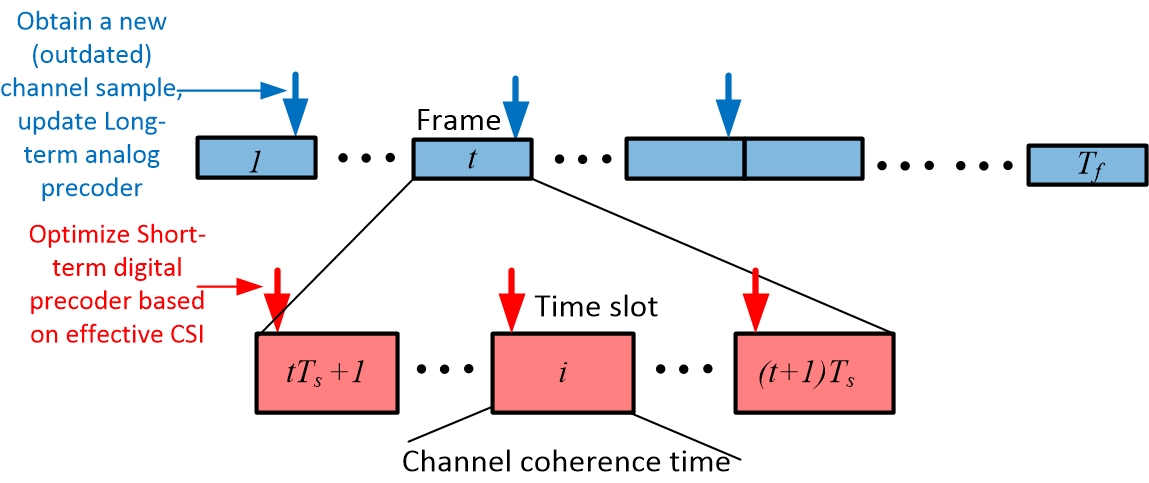}
\par\end{centering}
\caption{\label{fig:framestruc}An illustration of two-timescale frame structure. }
\end{figure}

In this paper, we focus on a coherence time interval of channel statistics
within which the channel statistics (distribution) are assumed to
be constant. The coherence time of channel statistics is divided into
$T_{f}$ frames and each frame consists of $T_{s}$ time slots, as
illustrated in Fig. \ref{fig:framestruc}. The channel state $\boldsymbol{H}=\left\{ \boldsymbol{H}_{n},\forall n\right\} $
is assumed to be constant within each time slot. In this paper, we
assume that one (possibly outdated) channel sample $\boldsymbol{H}$
at each frame can be obtained by uplink channel training. Specifically,
users send uplink pilot signals and then the BBU estimates the channel
based on the received pilot signals collected from RRHs via the fronthaul.
Several compressive sensing (CS) based channel estimation methods
have been proposed for uplink channel training with a limited number
of RF chains, see e.g., \cite{Liu_ICC2017_CRFCE,Lian_TSP18_WLASSO}.
At each time slot, the BBU needs to obtain the effective CSI $\boldsymbol{F}_{n}^{H}\boldsymbol{H}_{n}\in\mathbb{C}^{S\times K},\forall n$,
which can also be obtained by uplink channel training. Since the dimension
of the effective channel is equal to the number of RF chains at each
RRH, a simple least-square (LS) based channel estimation method is
sufficient to obtain a good estimation of the effective channel with
low computation time, i.e., the delay for effective CSI can be made
small relative to the channel coherence time. In our design, the BBU
is not required to have explicit knowledge of the channel statistics.
By observing one channel sample at each frame, the proposed algorithm
can automatically learn the channel statistics (in an implicit way).
Specifically, the long-term analog filtering matrices $\mathbf{F}_{n},\forall n$
are only updated once per frame based on a (possibly outdated) channel
sample to achieve massive MIMO array gain with reduced implementation
cost. On the other hand, the short-term control variables $\left\{ \boldsymbol{p},\boldsymbol{V}_{n},\boldsymbol{d}_{n},\boldsymbol{u}_{k}\right\} $
are adaptive to the real-time effective CSI $\boldsymbol{F}_{n}^{H}\boldsymbol{H}_{n},\forall n$
to achieve the spatial multiplexing gain. For convenience, we let
$\boldsymbol{v}=\textrm{Vec}\left(\left[\boldsymbol{V}_{1},...,\boldsymbol{V}_{N}\right]\right)$,
$\boldsymbol{d}=\left[\boldsymbol{d}_{1}^{T},...,\boldsymbol{d}_{N}^{T}\right]^{T}$
and $\boldsymbol{u}=\left[\boldsymbol{u}_{1}^{T},...,\boldsymbol{u}_{K}^{T}\right]^{T}$. 

For given long-term control variables $\boldsymbol{\theta}=\left[\boldsymbol{\theta}_{1}^{T},...,\boldsymbol{\theta}_{n}^{T}\right]^{T}$
(phase vectors of analog filtering matrices), short-term control variables
$\boldsymbol{x}\triangleq\left[\boldsymbol{p}^{T},\boldsymbol{v}^{T},\boldsymbol{d}^{T},\boldsymbol{u}^{T}\right]^{T}$
and channel realization $\boldsymbol{H}$, the achievable data rate
of user $k$ is given by
\[
r_{k}^{\circ}\left(\boldsymbol{\theta},\boldsymbol{x},\boldsymbol{H}\right)=\log\left(1+\text{SINR}{}_{k}\left(\boldsymbol{\theta},\boldsymbol{x};\boldsymbol{H}\right)\right),
\]
where $\text{SINR}{}_{k}\left(\boldsymbol{\theta},\boldsymbol{x};\boldsymbol{H}\right)$
is the SINR of user $k$ given by
\begin{align*}
\text{SINR}{}_{k}\left(\boldsymbol{\theta},\boldsymbol{x};\boldsymbol{H}\right)=\;\;\;\;\;\;\;\;\;\;\;\;\;\;\;\;\;\;\;\;\;\;\;\;\;\;\;\;\;\;\;\;\;\;\;\;\;\;\;\;\;\;\;\;\;\;\;\;\;\;\;\;\;\;\;\;\;\;\;\;\;\;\\
\frac{p_{k}|\boldsymbol{u}_{k}^{H}\widetilde{\boldsymbol{V}}^{H}\boldsymbol{h}_{k}|^{2}}{\underset{l\neq k}{\sum}p_{l}|\boldsymbol{u}_{k}^{H}\widetilde{\boldsymbol{V}}^{H}\boldsymbol{h}_{l}|^{2}+||\boldsymbol{u}_{k}^{H}\widetilde{\boldsymbol{V}}^{H}||^{2}+\boldsymbol{u}_{k}^{H}\boldsymbol{Q}\left(\boldsymbol{\theta},\boldsymbol{p},\boldsymbol{v},\boldsymbol{d}\right)\boldsymbol{u}_{k}},
\end{align*}
where
\begin{flalign*}
 & \boldsymbol{Q}\left(\boldsymbol{\theta},\boldsymbol{p},\boldsymbol{v},\boldsymbol{d}\right)=\\
 & \text{diag}\left(\boldsymbol{Q}_{1}\left(\boldsymbol{p},\boldsymbol{F}_{1}\boldsymbol{V}_{1},\boldsymbol{d}_{1}\right),...,\boldsymbol{Q}_{N}\left(\boldsymbol{p},\boldsymbol{F}_{N}\boldsymbol{V}_{N},\boldsymbol{d}_{N}\right)\right).
\end{flalign*}
Note that $\boldsymbol{F}_{n}$ is a function of $\boldsymbol{\theta}_{n}$
and we will explicitly write it as $\boldsymbol{F}_{n}\left(\boldsymbol{\theta}_{n}\right)$. 

Let $\boldsymbol{x}(\boldsymbol{H})$ denote the short-term control
variable under channel state $\boldsymbol{H}$ and $\widetilde{\Omega}\triangleq\left\{ \boldsymbol{x}(\boldsymbol{H})\in\widetilde{\mathcal{X}},\forall\boldsymbol{H}\right\} $
denote the collection of the short-term control variables for all
possible channel states, with $\widetilde{\mathcal{X}}$ denoting
the feasible set of the short-term control variables. To be more specific,
$\widetilde{\mathcal{X}}$ is the set of all short-term control variables
$\boldsymbol{x}=\left[\boldsymbol{p}^{T},\boldsymbol{v}^{T},\boldsymbol{d}^{T},\boldsymbol{u}^{T}\right]^{T}$
that satisfy the following constraints:
\begin{align}
p_{k}\in\left[0,P_{k}\right], & \forall k,\label{eq:Pwocons}\\
2B_{W}\sum_{l=1}^{L}d_{n,l}\leq C_{n}, & \forall n,\label{eq:frontcons}\\
d_{n,l}\geq0\text{ is an integer,} & \forall n,l,\label{eq:dcons}
\end{align}
where $P_{k}$ is the individual power constraint at user $k$, $B_{W}$
is the system bandwidth, and (\ref{eq:frontcons}) is the fronthaul
capacity constraint. Then the average data rate of user $k$ is
\[
\overline{r}_{k}^{\circ}\left(\boldsymbol{\theta},\widetilde{\Omega}\right)=\mathbb{E}\left[r_{k}^{\circ}\left(\boldsymbol{\theta},\boldsymbol{x}(\boldsymbol{H});\boldsymbol{H}\right)\right],
\]
where the expectation is taken with respect to the channel state $\boldsymbol{H}$.
For convenience, define $\overline{\boldsymbol{r}}^{\circ}\left(\boldsymbol{\theta},\widetilde{\Omega}\right)\triangleq[\overline{r}_{1}^{\circ}\left(\boldsymbol{\theta},\widetilde{\Omega}\right),...,\overline{r}_{K}^{\circ}\left(\boldsymbol{\theta},\widetilde{\Omega}\right)]^{T}$
as the average data rate vector.

\section{Two-timescale Joint Optimization at BBU\label{sec:Two-timescale-Joint-Optimization}}

\subsection{Problem Formulation}

Note that $r_{k}^{\circ}\left(\boldsymbol{\theta},\boldsymbol{x},\boldsymbol{H}\right)$
is not a continuous function of $d_{n,l},\forall n,l$ because $d_{n,l}$
is an integer. To make the problem tractable, we relax the integer
constraint on $d_{n,l}$ and approximate the quantization noise power
$q_{n,l},\forall n,l$ with the following continuous function of a
real variable $d_{n,l}\geq0$ as
\begin{equation}
\hat{q}_{n,l}=\frac{3}{4^{d_{n,l}}}(\sum_{k=1}^{K}p_{k}|\boldsymbol{h}_{n,k}^{H}\widetilde{\boldsymbol{v}}_{n,l}|^{2}+\left\Vert \widetilde{\boldsymbol{v}}_{n,l}\right\Vert ^{2}).\label{eq:qapprox}
\end{equation}
The same approximation has also been considered in \cite{Liu_TSP2015_SCFCRAN}.
We use $r_{k}\left(\boldsymbol{\theta},\boldsymbol{x};\boldsymbol{H}\right)$
to denote the approximate data rate of user $k$ obtained by replacing
$q_{n,l}$ in (\ref{eq:qeq}) with $\hat{q}_{n,l}$ in (\ref{eq:qapprox})
and the integer constraint in (\ref{eq:dcons}) with constraint $d_{n,l}\geq0$.
Moreover, define $\overline{\boldsymbol{r}}\left(\boldsymbol{\theta},\Omega\right)=[\overline{r}_{1}\left(\boldsymbol{\theta},\Omega\right),...,\overline{r}_{K}\left(\boldsymbol{\theta},\Omega\right)]^{T}$
as the approximate average data rate vector, where $\overline{r}_{k}\left(\boldsymbol{\theta},\Omega\right)=\mathbb{E}\left[r_{k}\left(\boldsymbol{\theta},\boldsymbol{x}(\boldsymbol{H});\boldsymbol{H}\right)\right]$
and $\Omega\triangleq\left\{ \boldsymbol{x}(\boldsymbol{H})\in\mathcal{X},\forall\boldsymbol{H}\right\} $
with $\mathcal{X}$ denoting the set of all short-term control variables
that satisfy constraint (\ref{eq:Pwocons}), (\ref{eq:frontcons})
and $d_{n,l}\geq0$. To simplify the notation, we drop the arguments
in $r_{k}\left(\boldsymbol{\theta},\boldsymbol{x},\boldsymbol{H}\right)$,
$\overline{\boldsymbol{r}}\left(\boldsymbol{\theta},\Omega\right)$,
and write them as $r_{k}$, $\overline{\boldsymbol{r}}$, when there
is no ambiguity.

With the above approximate rate, the two-timescale joint optimization
of long-term and short-term control variables can be formulated as
the following utility maximization problem
\begin{equation}
\mathcal{P}:\:\max_{\boldsymbol{\theta}\in\Theta,\Omega}\:g\left(\overline{\boldsymbol{r}}\left(\boldsymbol{\theta},\Omega\right)\right),\label{eq:mainP-1}
\end{equation}
where the utility function $g\left(\overline{\boldsymbol{r}}\right)$
is continuously differentiable (and possibly non-concave) function
of $\overline{\boldsymbol{r}}$, $\Theta\triangleq\left[0,2\pi\right]^{NMS}$
is the feasible set of $\boldsymbol{\theta}$. Moreover, $g\left(\overline{\boldsymbol{r}}\right)$
is non-decreasing with respect to $\overline{r}_{k},\forall k$ and
its derivative $\nabla_{\overline{\boldsymbol{r}}}g\left(\overline{\boldsymbol{r}}\right)$
with respect to $\overline{\boldsymbol{r}}$ is Lipschitz continuous.
This general utility function $g\left(\overline{\boldsymbol{r}}\right)$
includes many important network utilities as special cases, such as
average sum rate ($g\left(\overline{\boldsymbol{r}}\right)=\sum_{k=1}^{K}\overline{r}_{k}$)
and proportional fairness utility ($g\left(\overline{\boldsymbol{r}}\right)=\sum_{k=1}^{K}\log\left(\overline{r}_{k}+\varepsilon\right)$,
where $\varepsilon>0$ is a small number to avoid the singularity
at $\overline{r}_{k}=0$). 

\subsection{Stationary Solution of Problem $\mathcal{P}$}

Since Problem $\mathcal{P}$ is a two-timescale stochastic non-convex
problem, we focus on designing an efficient algorithm to find stationary
solutions of Problem $\mathcal{P}$, as defined below.
\begin{defn}
[Stationary solution of $\mathcal{P}$]\label{def:Stationary-solution}A
solution $\left(\boldsymbol{\theta}^{*},\Omega^{*}=\left\{ \boldsymbol{x}^{*}\left(\boldsymbol{H}\right)\in\mathcal{X},\forall\boldsymbol{H}\right\} \right)$
is called a stationary solution of Problem $\mathcal{P}$ if it satisfies
the following conditions: 
\begin{enumerate}
\item For every $\boldsymbol{H}$ outside a set of probability zero, 
\begin{equation}
\left(\boldsymbol{x}-\boldsymbol{x}^{*}(\boldsymbol{H})\right)^{T}\mathbf{J}_{\boldsymbol{x}}\left(\boldsymbol{\theta}^{*},\boldsymbol{x}^{*}(\boldsymbol{H});\boldsymbol{H}\right)\nabla_{\overline{\boldsymbol{r}}}g\left(\overline{\boldsymbol{r}}^{*}\right)\leq0,\label{eq:KKTs}
\end{equation}
$\forall\boldsymbol{x}\in\mathcal{X},$ where $\mathbf{J}_{\boldsymbol{x}}\left(\boldsymbol{\theta}^{*},\boldsymbol{x}^{*}(\boldsymbol{H});\boldsymbol{H}\right)$
is the Jacobian matrix\footnote{The Jacobian matrix of $\boldsymbol{r}\left(\boldsymbol{\theta},\boldsymbol{x};\boldsymbol{H}\right)$
with respect to $\boldsymbol{x}$ is defined as $\mathbf{J}_{r}\left(\boldsymbol{\theta},\boldsymbol{x};\boldsymbol{H}\right)=\left[\begin{array}{cccc}
\nabla_{\boldsymbol{x}}r_{1} & \nabla_{\boldsymbol{x}}r_{2} & \cdots & \nabla_{\boldsymbol{x}}r_{K}\end{array}\right]$, where $\nabla_{\boldsymbol{x}}r_{k}$ is the partial derivative
of $r_{k}\left(\boldsymbol{\theta},\boldsymbol{x};\boldsymbol{H}\right)$
with respect to $\boldsymbol{x}$.} of the (approximate) rate vector $\boldsymbol{r}\triangleq[r_{1},...,r_{K}]^{T}$
with respect to $\boldsymbol{x}$ at $\boldsymbol{\theta}=\boldsymbol{\theta}^{*}$
and $\boldsymbol{x}=\boldsymbol{x}^{*}(\boldsymbol{H})$, and $\nabla_{\overline{\boldsymbol{r}}}g\left(\overline{\boldsymbol{r}}^{*}\right)$
is the derivative of $g\left(\overline{\boldsymbol{r}}\right)$ at
$\overline{\boldsymbol{r}}=\overline{\boldsymbol{r}}^{*}\triangleq\overline{\boldsymbol{r}}\left(\boldsymbol{\theta}^{*},\Omega^{*}\right)$.
\item 
\begin{equation}
\left(\boldsymbol{\theta}-\boldsymbol{\theta}^{*}\right)^{T}\nabla_{\boldsymbol{\theta}}g\left(\overline{\boldsymbol{r}}^{*}\right)\leq0,\forall\boldsymbol{\theta}\in\Theta,\label{eq:fix2}
\end{equation}
where $\nabla_{\boldsymbol{\theta}}g\left(\overline{\boldsymbol{r}}^{*}\right)\triangleq\mathbb{E}[\mathbf{J}_{\boldsymbol{\theta}}\left(\boldsymbol{\theta}^{*},\boldsymbol{x}^{*}(\boldsymbol{H});\boldsymbol{H}\right)]\nabla_{\overline{\boldsymbol{r}}}g\left(\overline{\boldsymbol{r}}^{*}\right)$
is the partial derivative of $g\left(\overline{\boldsymbol{r}}\right)$
with respect to $\boldsymbol{\theta}$ at $\boldsymbol{\theta}=\boldsymbol{\theta}^{*}$
and $\Omega=\Omega^{*}$, $\mathbf{J}_{\boldsymbol{\theta}}\left(\boldsymbol{\theta}^{*},\boldsymbol{x}^{*}(\boldsymbol{H});\boldsymbol{H}\right)$
is the Jacobian matrix\footnote{The Jacobian matrix of $\boldsymbol{r}\left(\boldsymbol{\theta},\boldsymbol{x};\boldsymbol{H}\right)$
with respect to $\boldsymbol{\theta}$ is defined as $\mathbf{J}_{r}\left(\boldsymbol{\theta},\boldsymbol{x};\boldsymbol{H}\right)=\left[\begin{array}{cccc}
\nabla_{\boldsymbol{\theta}}r_{1} & \nabla_{\boldsymbol{\theta}}r_{2} & \cdots & \nabla_{\boldsymbol{\theta}}r_{K}\end{array}\right]$, where $\nabla_{\boldsymbol{\theta}}r_{k}$ is the partial derivative
of $r_{k}\left(\boldsymbol{\theta},\boldsymbol{x};\boldsymbol{H}\right)$
with respect to $\boldsymbol{\theta}$.} of the (approximate) rate vector $\boldsymbol{r}\left(\boldsymbol{\theta},\boldsymbol{x};\boldsymbol{H}\right)$
with respect to $\boldsymbol{\theta}$ at $\boldsymbol{\theta}=\boldsymbol{\theta}^{*}$
and $\boldsymbol{x}=\boldsymbol{x}^{*}(\boldsymbol{H})$.
\end{enumerate}
\end{defn}

In other words, a solution $\left(\boldsymbol{\theta}^{*},\Omega^{*}\right)$
is called a stationary solution of $\mathcal{P}$ if for fixed $\boldsymbol{\theta}^{*},\left\{ \boldsymbol{x}^{*}(\boldsymbol{H}^{'})\forall\boldsymbol{H}^{'}\neq\boldsymbol{H}\right\} $,
$\boldsymbol{x}^{*}(\boldsymbol{H})$ is a stationary point of $\mathcal{P}$
w.p.1., and for fixed $\Omega^{*}$, $\boldsymbol{\theta}^{*}$ is
a stationary point of $\mathcal{P}$. The stationary solution is a
natural extension of the stationary point for a deterministic optimization
problem. The global optimal solution must be a stationary solution.
However, the set of stationary solutions may also contain local optimal
solutions and a certain type of saddle points. When $\mathcal{P}$
is a two-timescale stochastic convex problem, a stationary solution
$\left(\boldsymbol{\theta}^{*},\Omega^{*}\right)$ is also a globally
optimal solution. 

Note that a stationary solution $\left(\boldsymbol{\theta}^{*},\Omega^{*}\right)$
of $\mathcal{P}$ may not satisfy all the integer constraints in (\ref{eq:dcons}).
To obtain an integer solution for the quantization bits allocation,
we use the same method as in \cite{Liu_TSP2015_SCFCRAN} to round
each $d_{n,l}^{*}$ to its nearby integer as follows.
\[
\hat{d}_{n,l}\left(\alpha_{n}\right)=\begin{cases}
\left\lfloor d_{n,l}^{*}\right\rfloor , & \text{if }d_{n,l}^{*}-\left\lfloor d_{n,l}^{*}\right\rfloor \leq\alpha_{n},\\
\left\lceil d_{n,l}^{*}\right\rceil , & \text{otherwise},
\end{cases}\forall n,l,
\]
where $0\leq\alpha_{n}\leq1$ is chosen such that $\sum_{s=1}^{L}\hat{d}_{n,l}\left(\alpha_{n}\right)=C_{n}/2B$.
Since $\sum_{l=1}^{L}\hat{d}_{n,l}\left(1\right)=\sum_{l=1}^{L}\left\lfloor d_{n,l}^{*}\right\rfloor \leq C_{n}/2B$
and $\sum_{s=1}^{L}\hat{d}_{n,l}\left(0\right)=\left\lceil d_{n,l}^{*}\right\rceil \geq C_{n}/2B$,
we can always find such $\alpha_{n}$ using a bisection search over
$\alpha_{n}\in\left[0,1\right]$.

\section{Online Block-Coordinate Stochastic Successive Convex Approximation\label{sec:Constrained-Stochastic-Successiv}}

There are several challenges in finding stationary solutions of Problem
$\mathcal{P}$, elaborated as follows. 
\begin{center}
\fbox{\begin{minipage}[t]{0.96\columnwidth}%
\begin{challenge}\label{chl:Deterministic-Restriction}Complex coupling
between the short-term and long-term control variables; no closed-form
characterization of the average data rates $\overline{r}_{k}\left(\boldsymbol{\theta},\Omega\right),\forall k$;
unknown distribution of $\boldsymbol{H}$.\end{challenge}\vspace{-5bp}
\end{minipage}}
\par\end{center}

To the best of our knowledge, there lacks an efficient and online
algorithm with self-learning capability to handle the two-timescale
stochastic non-convex optimization problem $\mathcal{P}$. In this
section, we propose an online BC-SSCA algorithm to find stationary
solutions of Problem $\mathcal{P}$. We shall first summarize the
proposed BC-SSCA algorithm. Then we elaborate the implementation details.

\subsection{Summary of the BC-SSCA Algorithm}

The proposed online BC-SSCA algorithm is summarized in Algorithm 1
and its time line is illustrated in Fig. \ref{fig:framestruc}. In
BC-SSCA, an auxiliary weight vector $\boldsymbol{\mu}=\left[\mu_{1},...,\mu_{K}\right]^{T}$
is introduced to approximate the derivative $\nabla_{\overline{\boldsymbol{r}}}g\left(\overline{\boldsymbol{r}}\right)$.
At the beginning of each coherence time of channel statistics, the
BBU resets the BC-SSCA algorithm with an initial analog filter phase
vector $\boldsymbol{\theta}^{0}$ and a weight vector $\boldsymbol{\mu}^{0}$.
Then $\boldsymbol{\theta}$ and $\boldsymbol{\mu}$ are updated once
at the end of each frame, where $\boldsymbol{\theta}$ is updated
by maximizing a concave surrogate function $\bar{f}^{t}\left(\boldsymbol{\theta}\right)$
of $g\left(\overline{\boldsymbol{r}}\right)$ with respect to $\boldsymbol{\theta}$.
Note that we cannot obtain the optimal $\boldsymbol{\theta}$ by directly
maximizing $g\left(\overline{\boldsymbol{r}}\right)$ because $g\left(\overline{\boldsymbol{r}}\right)$
is not concave and it does not have closed-form expression. Specifically,
let $\boldsymbol{\theta}^{t}$ and $\boldsymbol{\mu}^{t}$ denote
the analog filter phase vector and weight vector used during the $t$-th
frame. The $t$-th iteration ($t$-th frame) of the BC-SSCA algorithm
is described as follows.

\subsubsection*{Step 1 (Short-term control optimization at each time slot)}

At time slot $i\in\left[tT_{s}+1,\left(t+1\right)T_{s}\right]$ in
the $t$-th frame, the BBU first acquires the effective channel $\boldsymbol{F}_{n}^{H}\left(\boldsymbol{\theta}^{t}\right)\boldsymbol{H}_{n}\left(i\right),\forall n$,
where $\boldsymbol{H}_{n}\left(i\right)$ is the channel state of
RRH $n$ at time slot $i$. Then it calculates the short-term control
variables $\boldsymbol{x}^{J_{t}}\left(\boldsymbol{\mu}^{t},\boldsymbol{\theta}^{t},\boldsymbol{H}(i)\right)$
from $\boldsymbol{F}_{n}^{H}\left(\boldsymbol{\theta}^{t}\right)\boldsymbol{H}_{n}(i),\forall n$
by running a \textit{short-term block-coordinate (BC) algorithm} with
input $J_{t}$, $\boldsymbol{\mu}^{t},\boldsymbol{\theta}^{t}$ and
$\boldsymbol{H}_{n}(i)$, where $J_{t}$ determines the total number
of iterations for the short-term BC algorithm at frame $t$. Note
that $\boldsymbol{x}^{J_{t}}\left(\boldsymbol{\mu}^{t},\boldsymbol{\theta}^{t},\boldsymbol{H}(i)\right)$
depends on $\boldsymbol{\theta}^{t},\boldsymbol{H}(i)$ only through
the effective channel $\boldsymbol{F}_{n}^{H}\left(\boldsymbol{\theta}^{t}\right)\boldsymbol{H}_{n}(i),\forall n$. 

Specifically, for given input $J$, $\boldsymbol{\mu},\boldsymbol{\theta}$
and $\boldsymbol{H}$, the short-term BC algorithm runs $J$ iterations
to find a stationary point (up to certain accuracy) of the following
weighted sum-rate maximization problem (WSRMP): 
\[
\mathcal{P}_{S}\left(\boldsymbol{\mu},\boldsymbol{\theta},\boldsymbol{H}\right):\:\max_{\boldsymbol{x}=\left[\boldsymbol{p}^{T},\boldsymbol{v}^{T},\boldsymbol{d}^{T},\boldsymbol{u}^{T}\right]^{T}}\sum_{k=1}^{K}\mu_{k}r_{k}\left(\boldsymbol{\theta},\boldsymbol{x};\boldsymbol{H}\right).
\]
The reason that the short-term control variables $\boldsymbol{x}^{J_{t}}\left(\boldsymbol{\mu}^{t},\boldsymbol{\theta}^{t},\boldsymbol{H}(i)\right)$
are obtained by solving the WSRMP $\mathcal{P}_{S}\left(\boldsymbol{\mu}^{t},\boldsymbol{\theta}^{t},\boldsymbol{H}(i)\right)$
is as follows. It follows from (\ref{eq:KKTs}) that at a stationary
solution $\left(\boldsymbol{\theta}^{*},\Omega^{*}=\left\{ \boldsymbol{x}^{*}\left(\boldsymbol{H}\right)\in\mathcal{X},\forall\boldsymbol{H}\right\} \right)$,
the short-term control variables $\boldsymbol{x}^{*}(\boldsymbol{H})$
for channel realization $\boldsymbol{H}$ must be a stationary point
of $\mathcal{P}_{S}\left(\boldsymbol{\mu}^{*},\boldsymbol{\theta}^{*},\boldsymbol{H}\right)$
with a \textit{stationary weight vector} $\boldsymbol{\mu}^{*}=\nabla_{\overline{\boldsymbol{r}}}g\left(\overline{\boldsymbol{r}}^{\ast}\right)$.
Therefore, for fixed long-term control variable $\boldsymbol{\theta}^{*}$,
once we know $\boldsymbol{\mu}^{*}$, the joint optimization of the
collection of short-term control variables $\Omega$ can be decoupled
into the optimization of per time slot short-term control variables
by solving a WSRMP $\mathcal{P}_{S}\left(\boldsymbol{\mu}^{*},\boldsymbol{\theta}^{*},\boldsymbol{H}\left(i\right)\right)$
at each time slot $i$. However, $\boldsymbol{\mu}^{*}$ is not known
a prior. Therefore, the basic idea of the proposed algorithm is to
iteratively update the long-term variable $\boldsymbol{\theta}^{t}$
and the weight vector $\boldsymbol{\mu}^{t}$ such that $\boldsymbol{\theta}^{t}$
and $\boldsymbol{\mu}^{t}$ converge to a stationary solution $\boldsymbol{\theta}^{*}$
and the corresponding stationary weight vector $\boldsymbol{\mu}^{*}$,
respectively. Then the short-term control variable $\boldsymbol{x}^{*}(\boldsymbol{H})$
that satisfies (\ref{eq:KKTs}) for each channel state $\boldsymbol{H}$
can be calculated by finding a stationary point of the corresponding
WSRMP $\mathcal{P}_{S}\left(\boldsymbol{\mu}^{t},\boldsymbol{\theta}^{t},\boldsymbol{H}\right)$
as $t\rightarrow\infty$.

The details of the short-term BC algorithm will be postponed to Section
\ref{subsec:short-term algorithm}. Here, we only discuss the impact
of $J_{t}$ on the convergence. For any finite iteration $t<\infty$,
$J_{t}$ is finite, and we can let $J_{t}\rightarrow\infty$ as $t\rightarrow\infty$
to ensure the convergence to stationary solutions. A larger $J_{t}$
for fixed $t$ usually leads to a faster overall convergence speed
at the cost of higher complexity. 

\subsubsection*{Step 2 (Long-term control optimization at the end of frame $t$)}

In Step 2a, the BBU obtains a full channel sample $\boldsymbol{H}^{t}\triangleq\boldsymbol{H}(tT_{s}+1)$
before the end of $t$-th frame. Then, in Step 2b (at the end of the
$t$-th frame), the BBU updates the surrogate function $\bar{f}^{t}\left(\boldsymbol{\theta}\right)$
based on $\boldsymbol{H}^{t}$, the current iterate $\boldsymbol{\theta}^{t}$,
and the short-term control variables $\boldsymbol{x}\left(i\right)\triangleq\boldsymbol{x}^{J_{t}}\left(\boldsymbol{\mu}^{t},\boldsymbol{\theta}^{t},\boldsymbol{H}(i)\right),\forall i\in\left[tT_{s}+1,\left(t+1\right)T_{s}\right]$
as
\begin{align}
\bar{f}^{t}\left(\boldsymbol{\theta}\right) & =g\left(\hat{\boldsymbol{r}}^{t}\right)+\left(\mathbf{f}^{t}\right)^{T}\left(\boldsymbol{\theta}-\boldsymbol{\theta}^{t}\right)-\tau\left\Vert \boldsymbol{\theta}-\boldsymbol{\theta}^{t}\right\Vert ^{2},\label{eq:upsurrgate}
\end{align}
where $\tau>0$ is a constant; $\hat{\boldsymbol{r}}^{t}=\left[\hat{r}_{1}^{t},...,\hat{r}_{K}^{t}\right]^{T}$
is an approximation for the average data rate vector, which is updated
recursively as
\begin{equation}
\hat{r}_{k}^{t}=\left(1-\rho_{t}\right)\hat{r}_{k}^{t-1}+\rho_{t}\sum_{i=tT_{s}+1}^{\left(t+1\right)T_{s}}\frac{r_{k}\left(\boldsymbol{\theta}^{t},\boldsymbol{x}\left(i\right);\boldsymbol{H}\left(i\right)\right)}{T_{s}},\forall k,\label{eq:fit}
\end{equation}
with $\hat{r}_{k}^{-1}=0,\forall k$; $\mathbf{f}^{t}$ is an approximation
of the partial derivative $\nabla_{\boldsymbol{\theta}}g\left(\overline{\boldsymbol{r}}\left(\boldsymbol{\theta},\Omega\right)\right)$
with respect to $\boldsymbol{\theta}$, which is updated recursively
as
\begin{align}
\mathbf{F}^{t} & =\left(1-\rho_{t}\right)\mathbf{F}^{t-1}+\rho_{t}\mathbf{J}_{\boldsymbol{\theta}}\left(\boldsymbol{\theta}^{t},\boldsymbol{x}\left(tT_{s}+1\right);\boldsymbol{H}^{t}\right),\nonumber \\
\mathbf{f}^{t} & =\mathbf{F}^{t}\nabla_{\overline{\boldsymbol{r}}}g\left(\hat{\boldsymbol{r}}^{t}\right),\label{eq:ft}
\end{align}
with $\mathbf{F}^{-1}=\boldsymbol{0}$, where $\rho_{t}\in\left(0,1\right]$
is a sequence to be properly chosen, $\mathbf{J}_{\boldsymbol{\theta}}\left(\boldsymbol{\theta},\boldsymbol{x};\boldsymbol{H}\right)$
is the Jacobian matrix of the rate vector $\boldsymbol{r}\left(\boldsymbol{\theta},\boldsymbol{x};\boldsymbol{H}\right)$
with respect to $\boldsymbol{\theta}$ and its expression is derived
in Appendix \ref{subsec:Jacobian-Matrix-of}, $\mathbf{F}^{t}$ is
an approximation for $\mathbb{E}\left[\mathbf{J}_{\boldsymbol{\theta}}\left(\boldsymbol{\theta}^{t},\boldsymbol{x}^{J}\left(\boldsymbol{\mu}^{t},\boldsymbol{\theta}^{t},\boldsymbol{H}\right);\boldsymbol{H}\right)\right]$.
It will be shown in Lemma \ref{lem:Convergence-of-Surro} that $\hat{r}_{k}^{t}$
and $\mathbf{f}^{t}$ will converge to the true average data rate
and partial derivative, respectively. Therefore, the issues of no
closed-form characterization of the average data rates $\overline{r}_{k}\left(\boldsymbol{\theta},\Omega\right),\forall k$
and unknown distribution of $\boldsymbol{H}$ can be addressed by
approximating the average data rate and $\nabla_{\boldsymbol{\theta}}g\left(\overline{\boldsymbol{r}}\left(\boldsymbol{\theta},\Omega\right)\right)$
in a recursive way as in (\ref{eq:fit}) and (\ref{eq:ft}) based
on the online observations of the channel samples $\boldsymbol{H}(i)$
at each time slot $i$. Moreover, the weight vector $\boldsymbol{\mu}$
is updated as 
\begin{equation}
\boldsymbol{\mu}^{t+1}=\left(1-\gamma_{t}\right)\boldsymbol{\mu}^{t}+\gamma_{t}\bar{\boldsymbol{\mu}}^{t}.\label{eq:updatemu}
\end{equation}
with $\bar{\boldsymbol{\mu}}^{t}\triangleq\nabla_{\overline{\boldsymbol{r}}}g\left(\hat{\boldsymbol{r}}^{t}\right)$,
where $\gamma_{t}\in\left(0,1\right]$ is a sequence satisfying $\sum_{t}\gamma_{t}=\infty$,
$\sum_{t}\left(\gamma_{t}\right)^{2}<\infty$.

In Step 2c, the optimal solution $\bar{\boldsymbol{\theta}}^{t}$
of the following quadratic optimization problem is solved:
\begin{align}
\bar{\boldsymbol{\theta}}^{t}=\underset{\boldsymbol{\theta}\in\Theta}{\text{argmax}}\: & \bar{f}^{t}\left(\boldsymbol{\theta}\right),\label{eq:Pitert}
\end{align}
which has closed-form solution $\bar{\boldsymbol{\theta}}^{t}=\mathbb{P}_{\Theta}\left[\boldsymbol{\theta}^{t}+\frac{\mathbf{f}^{t}}{2\tau}\right]$,
where $\mathbb{P}_{\Theta}\left[\cdot\right]$ denotes the projection
on to the box feasible region $\Theta$. Finally, $\boldsymbol{\theta}$
is updated according to
\begin{equation}
\boldsymbol{\theta}^{t+1}=\left(1-\gamma_{t}\right)\boldsymbol{\theta}^{t}+\gamma_{t}\bar{\boldsymbol{\theta}}^{t}.\label{eq:updatext}
\end{equation}
Then the above iteration is carried out until convergence.

\begin{algorithm}
\caption{\label{alg1}Block-Coordinate Stochastic Successive Convex Approximation}

\textbf{\small{}Input: }{\small{}$\left\{ \rho^{t}\right\} $, $\left\{ \gamma^{t}\right\} $,
$\left\{ J_{t}\right\} $.}{\small\par}

\textbf{\small{}Initialize:}{\small{} $\boldsymbol{\theta}^{0}\in\Theta$;
$\boldsymbol{\mu}^{0}=\left[1,...,1\right]^{T}$, $t=0$.}{\small\par}

\textbf{\small{}Step 1 }{\small{}(}\textbf{\small{}Short-term control
optimization at each time slot}{\small{} $i\in\left[tT_{s}+1,\left(t+1\right)T_{s}\right]$):}{\small\par}

\textbf{\small{}\,\,\,\,\,\,\,\,\,\,}\textit{\small{}Apply
the}\textbf{\textit{\small{} }}\textit{\small{}short-term BC algorithm
with input $J_{t}$, $\boldsymbol{\mu}^{t},\boldsymbol{\theta}^{t}$
and $\boldsymbol{H}_{n}(i)$, to obtain the short-term variable $\boldsymbol{x}^{J_{t}}\left(\boldsymbol{\mu}^{t},\boldsymbol{\theta}^{t},\boldsymbol{H}(i)\right)$,
as elaborated in Section \ref{subsec:short-term algorithm}.}{\small\par}

\textbf{\small{}Step 2 }{\small{}(}\textbf{\small{}Long-term control
optimization at the end of frame}{\small{} $t$): }{\small\par}

\textbf{\small{}\,\,\,\,\,\,\,\,\,\,2a}{\small{}: }\textit{\small{}Obtain
a full channel sample $\boldsymbol{H}^{t}\triangleq\boldsymbol{H}(tT_{s}+1)$.}{\small\par}

\textbf{\small{}\,\,\,\,\,\,\,\,\,\,2b}{\small{}: }\textit{\small{}Update
the surrogate function $\bar{f}^{t}\left(\boldsymbol{\theta}\right)$
according to (\ref{eq:upsurrgate}) based on $\boldsymbol{H}^{t},$$\boldsymbol{\theta}^{t}$
and $\boldsymbol{x}^{J_{t}}\left(\boldsymbol{\mu}^{t},\boldsymbol{\theta}^{t},\boldsymbol{H}(i)\right),\forall i\in\left[tT_{s}+1,\left(t+1\right)T_{s}\right]$.
Calculate $\bar{\boldsymbol{\mu}}^{t}=\nabla_{\overline{\boldsymbol{r}}}g\left(\hat{\boldsymbol{r}}^{t}\right)$
and update $\boldsymbol{\mu}^{t+1}$ according to (\ref{eq:updatemu}).}{\small\par}

\textbf{\small{}\,\,\,\,\,\,\,\,\,\,2c}{\small{}: }\textit{\small{}Solve
(\ref{eq:Pitert}) to obtain $\bar{\boldsymbol{\theta}}^{t}$.}\textbf{\textit{\small{}
}}\textit{\small{}Update $\boldsymbol{\theta}^{t+1}$ according to
(\ref{eq:updatext}).}{\small\par}

\textbf{\small{}Let}{\small{} $t=t+1$ and return to Step 1.}{\small\par}
\end{algorithm}

\subsection{Short-term Block-Coordinate Algorithm\label{subsec:short-term algorithm}}

To apply the BC algorithm, we first transform the WSRMP $\mathcal{P}_{S}\left(\boldsymbol{\mu},\boldsymbol{\theta},\boldsymbol{H}\right)$
to the following weighted minimum mean square error (WMMSE) problem
\begin{align}
\min_{\boldsymbol{\beta},\boldsymbol{v},\boldsymbol{d},\boldsymbol{u},\boldsymbol{w}} & \sum_{k=1}^{K}\mu_{k}\left(w_{k}\eta_{k}-\textrm{log}w_{k}\right)\label{eq:WMMSE}\\
\text{s.t. } & \boldsymbol{d}\geq\boldsymbol{0},\text{ (\ref{eq:Pwocons}) and (\ref{eq:frontcons})},\nonumber 
\end{align}
where $\boldsymbol{w}=\left[w_{1},...,w_{K}\right]$ with $w_{k}>0:\forall k$
is a weight vector for MSE, $\boldsymbol{\beta}=\left[\beta_{1},...,\beta_{k}\right]^{T}$
with $\left|\beta_{k}\right|^{2}=p_{k}$ and 
\begin{align*}
 & \eta_{k}\triangleq\mathbb{E}\left[\left|s_{k}-\hat{s}_{k}\right|^{2}|\boldsymbol{H}\right]\\
= & \left|1-\boldsymbol{u}_{k}^{H}\widetilde{\boldsymbol{V}}^{H}\boldsymbol{h}_{k}\beta_{k}\right|^{2}+\sum_{l\neq k}\left|\boldsymbol{u}_{k}^{H}\widetilde{\boldsymbol{V}}^{H}\boldsymbol{h}_{l}\beta_{l}\right|^{2}\\
+ & \boldsymbol{u}_{k}^{H}\widetilde{\boldsymbol{V}}^{H}\widetilde{\boldsymbol{V}}\boldsymbol{u}_{k}+\boldsymbol{u}_{k}^{H}\boldsymbol{Q}\left(\boldsymbol{\theta},\boldsymbol{p},\boldsymbol{v},\boldsymbol{d}\right)\boldsymbol{u}_{k},
\end{align*}
is the MSE of user $k$. Following similar proof to that of Theorem
1 in \cite{Luo_TSP11_WMMSE}, it can be shown that Problem $\mathcal{P}_{S}\left(\boldsymbol{\mu},\boldsymbol{\theta},\boldsymbol{H}\right)$
is equivalent to (\ref{eq:WMMSE}). Moreover, if $\left(\boldsymbol{\beta}^{*},\boldsymbol{v}^{*},\boldsymbol{d}^{*},\boldsymbol{u}^{*},\boldsymbol{w}^{*}\right)$
is a stationary point of (\ref{eq:WMMSE}), then $\left(\boldsymbol{p}^{*},\boldsymbol{v}^{*},\boldsymbol{d}^{*},\boldsymbol{u}^{*}\right)$
is also a stationary point of $\mathcal{P}_{S}\left(\boldsymbol{\mu},\boldsymbol{\theta},\boldsymbol{H}\right)$,
where $\boldsymbol{p}^{*}=\left[p_{1}^{*},...,p_{K}^{*}\right]^{T}$
with $p_{k}^{*}=\left|\beta_{k}^{*}\right|^{2}$. Therefore, we shall
focus on designing a BC algorithm to find a stationary point of (\ref{eq:WMMSE}).

In the proposed BC algorithm, the short-term control variables $\boldsymbol{\beta},\boldsymbol{v},\boldsymbol{d},\boldsymbol{u},\boldsymbol{w}$
are optimized in an alternating way by solving a convex subproblem
with respect to each variable. The BC algorithm is summarized in Algorithm
2. The choice of the initial point and the update equation for each
variable is elaborated below.

\begin{algorithm}
\caption{Short-term Block-Coordinate Algorithm for $\mathcal{P}_{S}\left(\boldsymbol{\mu},\boldsymbol{\theta},\boldsymbol{H}\right)$}

\textbf{\small{}Input:}{\small{} $J$, $\boldsymbol{\mu},\boldsymbol{\theta}$
and $\boldsymbol{H}$.}{\small\par}

\textbf{\small{}Initialization:}{\small{} }\textbf{\small{}Let}{\small{}
$j=0$, $\beta_{k}=\sqrt{P_{k}},\forall k$, $d_{n,l}=\frac{C_{n}}{2BL},\forall n,l$
and $\boldsymbol{V}_{n},\forall n$ be the first $L$ eigenvectors
of $\boldsymbol{F}_{n}^{H}\boldsymbol{H}_{n}\boldsymbol{H}_{n}^{H}\boldsymbol{F}_{n}$.}{\small\par}

\textbf{\small{}Step 1 (Update }{\small{}$\boldsymbol{u}$, $\boldsymbol{w}$
and $\boldsymbol{\beta}$}\textbf{\small{}): For $k=1,...,K$, let
\begin{equation}
\boldsymbol{u}_{k}=\left(\sum_{l=1}^{K}\widetilde{\boldsymbol{V}}^{H}\boldsymbol{h}_{l}\left|\beta_{l}\right|^{2}\boldsymbol{h}_{l}^{H}\widetilde{\boldsymbol{V}}+\widetilde{\boldsymbol{V}}^{H}\widetilde{\boldsymbol{V}}+\boldsymbol{Q}\right)^{-1}\widetilde{\boldsymbol{V}}^{H}\boldsymbol{h}_{k}\beta_{k},\label{eq:mmse}
\end{equation}
\begin{equation}
w_{k}=\left(1-\boldsymbol{u}_{k}^{H}\widetilde{\boldsymbol{V}}^{H}\boldsymbol{h}_{k}\beta_{k}\right)^{-1},\label{eq:upw}
\end{equation}
}{\small{}
\begin{equation}
\beta_{k}=\beta_{k}^{*}\left(\lambda_{k}\right),\label{eq:upbeta}
\end{equation}
where $\beta_{k}^{*}\left(\lambda_{k}\right)$ is given in (\ref{eq:upbeta1}).}{\small\par}

\textbf{\small{}Step 2 (Update }{\small{}$\boldsymbol{v}$}\textbf{\small{}):
Let }{\small{}$\boldsymbol{v}^{'}=\boldsymbol{v}$.}\textbf{\small{}
}{\small{}Update $\boldsymbol{v}$ according to (\ref{eq:optv-1}),
which depends on $\boldsymbol{v}^{'}$.}{\small\par}

\textbf{\small{}Step 3 (Update $\boldsymbol{d}$): Let }{\small{}$d_{n,l}=d_{n,l}^{*}\left(\lambda_{n}\right),\forall n,l$,
where $d_{n,l}^{*}\left(\lambda_{n}\right)$ is given in (\ref{eq:optd}).}{\small\par}

\textbf{\small{}Let}{\small{} $j=j+1$. If $j=J$,}\textbf{\small{}
terminate }{\small{}the algorithm and}\textbf{\small{} output}{\small{}
$\boldsymbol{x}^{J}\left(\boldsymbol{\mu},\boldsymbol{\theta},\boldsymbol{H}\right)=\left[\boldsymbol{p}^{T},\boldsymbol{v}^{T},\boldsymbol{d}^{T},\boldsymbol{u}^{T}\right]^{T}$,
where $p_{k}=\left|\beta_{k}\right|^{2},\forall k$. Otherwise, }\textbf{\small{}go
to}{\small{} Step 1. }{\small\par}
\end{algorithm}

\subsubsection{Choice of Initial Point}

For $\boldsymbol{\beta}$, we choose the initial point to be $\beta_{k}=\sqrt{P_{k}},\forall k$,
i.e., each user transmits at the maximum power. For $\boldsymbol{d}$,
we choose the initial point to be $d_{n,l}=\frac{C_{n}}{2BL},\forall n,l$,
i.e., equal quantization bits allocation at each RRH. For $\boldsymbol{v}$,
we choose $\boldsymbol{V}_{n},\forall n$ to be the first $L$ eigenvectors
of $\boldsymbol{F}_{n}^{H}\boldsymbol{H}_{n}\boldsymbol{H}_{n}^{H}\boldsymbol{F}_{n}$.

\subsubsection{Optimization of $\boldsymbol{u}$, $\boldsymbol{w}$ and $\boldsymbol{\beta}$}

When fixing the other short-term variables, the optimal $\boldsymbol{u}$
is given by the MMSE receiver in (\ref{eq:mmse}), where $\boldsymbol{Q}$
is an abbreviation for $\boldsymbol{Q}\left(\boldsymbol{\theta},\boldsymbol{p},\boldsymbol{v},\boldsymbol{d}\right)$;
the optimal $w_{k}$ is given by (\ref{eq:upw}); and the optimal
$\boldsymbol{\beta}$ is given by $\beta_{k}=\beta_{k}^{*}\left(\lambda_{k}\right),\forall k$
with
\begin{align}
\beta_{k}^{*}\left(\lambda_{k}\right) & =\mu_{k}w_{k}\textrm{Re}\left[\boldsymbol{u}_{k}^{H}\widetilde{\boldsymbol{V}}^{H}\boldsymbol{h}_{k}\right]\nonumber \\
 & \times\left(\sum_{l=1}^{K}2\mu_{l}w_{l}\boldsymbol{h}_{k}^{H}\widetilde{\boldsymbol{V}}\boldsymbol{u}_{l}\boldsymbol{u}_{l}^{H}\widetilde{\boldsymbol{V}}^{H}\boldsymbol{h}_{k}+\nu_{k}+2\lambda_{k}\right)^{-1},\label{eq:upbeta1}
\end{align}
where $\nu_{k}=\sum_{n,l}\frac{6}{4^{d_{n,l}}}\left|u_{k,n,l}\right|^{2}|\boldsymbol{h}_{n,k}^{H}\widetilde{\boldsymbol{v}}_{n,l}|^{2}$,
$u_{k,n,l}$ is the $\left(\left(n-1\right)N+s\right)$-th element
of $\boldsymbol{u}_{k}$, and $\lambda_{k}$ is chosen to be zero
if $\left|\beta_{k}^{*}\left(0\right)\right|^{2}\leq P_{k}$ and chosen
to satisfy $\left|\beta_{k}^{*}\left(\lambda_{k}\right)\right|^{2}=P_{k}$
otherwise.

\subsubsection{Optimization of $\boldsymbol{v}$}

When fixing the other short-term variables, the optimization of $\boldsymbol{v}$
is not necessarily a strictly convex problem and the optimal $\boldsymbol{v}$
may not be unique. To ensure the convergence of the short-term BC
algorithm, we solve the following modified subproblem with respect
to $\boldsymbol{v}$ by adding a proximal regularization term $\epsilon\left\Vert \boldsymbol{v}-\boldsymbol{v}^{'}\right\Vert ^{2}$:
\begin{equation}
\min_{\boldsymbol{v}}\sum_{k=1}^{K}\mu_{k}\left(w_{k}\eta_{k}-\textrm{log}w_{k}\right)+\epsilon\left\Vert \boldsymbol{v}-\boldsymbol{v}^{'}\right\Vert ^{2},\label{eq:WMMSE-1}
\end{equation}
where $\boldsymbol{v}^{'}$ is the digital filter at the beginning
of the current iteration in Algorithm 2, and $\epsilon>0$ is a small
positive number.

Clearly, Problem (\ref{eq:WMMSE-1}) is an unconstrained quadratic
optimization problem. Therefore, we can obtain the optimal digital
filter by checking its first-order optimality condition. After some
tedious calculations, it can be shown that the first-order optimality
condition can be expressed in a compact form as

\begin{equation}
\bm{B}v+\bm{J}+\epsilon(\boldsymbol{v}-\boldsymbol{v}^{'})=0,\label{eq:firstopt}
\end{equation}
where the LHS is the gradient of the objective function in (\ref{eq:WMMSE-1}),
$\boldsymbol{B}=\left[\boldsymbol{B}_{1,1}^{T},...,\boldsymbol{B}_{N,L}^{T}\right]^{T}$
and $\boldsymbol{B}_{n,l}=[\boldsymbol{B}_{1,1,n,l},...,\boldsymbol{B}_{N,L,n,l}]^{T}$
with
\begin{align*}
 & \boldsymbol{B}_{n^{'},l^{'},n,l}\\
= & \begin{cases}
\sum_{k=1}^{K}\mu_{k}w_{k}|u_{k,n,l}|^{2}(\frac{3}{4^{d_{n,l}}}+1)\boldsymbol{D}_{n}, & n^{'}=n,l^{'}=l,\\
\sum_{k=1}^{K}\mu_{k}w_{k}u_{k,n,l}^{\ast}u_{k,n,l^{'}}\boldsymbol{D}_{n}, & n^{'}=n,l^{'}\neq l,\\
\sum_{k=1}^{K}\mu_{k}w_{k}u_{k,n,l}^{\ast}u_{k,n^{'},l^{'}}\boldsymbol{D}_{n,n^{'}}, & n^{'}\neq n,
\end{cases}\\
\boldsymbol{D}_{n} & =\boldsymbol{F}_{n}^{H}\boldsymbol{F}_{n}+\sum_{k=1}^{K}\beta_{k}^{2}\boldsymbol{F}_{n}^{H}\boldsymbol{h}_{n,k}\boldsymbol{h}_{n,k}^{H}\boldsymbol{F}_{n},\\
\boldsymbol{D}_{n,n^{'}} & =\sum_{k=1}^{K}\beta_{k}^{2}\boldsymbol{F}_{n}^{H}\boldsymbol{h}_{n,k}\boldsymbol{h}_{n^{'},k}\boldsymbol{F}_{n},
\end{align*}
and $\boldsymbol{J}=\left[\boldsymbol{J}_{1,1},...,\boldsymbol{J}_{N,L}\right]^{T}$
with $\boldsymbol{J}_{n,l}=\sum_{k=1}^{K}\mu_{k}w_{k}\beta_{k}u_{k,n,l}^{\ast}\boldsymbol{F}_{n}^{H}\boldsymbol{h}_{n,k},\forall n,l$.
From (\ref{eq:firstopt}), the optimal digital filter for (\ref{eq:WMMSE-1})
is given by
\begin{equation}
\boldsymbol{v}=(\boldsymbol{B}+\epsilon\boldsymbol{I})^{-1}(\boldsymbol{J}+\epsilon\boldsymbol{v}^{'}).\label{eq:optv-1}
\end{equation}

\subsubsection{Optimization of $\boldsymbol{d}$}

The subproblem with respect to $\boldsymbol{d}$ can be expressed
as:
\begin{equation}
\min_{\boldsymbol{d}\geq\boldsymbol{0}}\sum_{k=1}^{K}\mu_{k}w_{k}\boldsymbol{u}_{k}^{H}\boldsymbol{Q}\boldsymbol{u}_{k},\text{ s.t. }\text{(\ref{eq:frontcons})},\label{eq:subpd}
\end{equation}
Note that we have
\begin{align*}
\boldsymbol{u}_{k}^{H}\boldsymbol{Q}\boldsymbol{u}_{k} & =\sum_{n=1}^{N}\boldsymbol{u}_{k,n}^{H}\boldsymbol{Q}_{n}\boldsymbol{u}_{k,n}=\sum_{n=1}^{N}\sum_{l=1}^{L}q_{n,l}|u_{k,n,l}|^{2}\\
 & =\sum_{n=1}^{N}\sum_{l=1}^{L}\varsigma_{k,n,l}4^{-d_{n,l}},\forall k,
\end{align*}
where $\boldsymbol{Q}_{n}$ is an abbreviation for $\boldsymbol{Q}_{n}\left(\boldsymbol{p},\boldsymbol{F}_{n}\boldsymbol{V}_{n},\boldsymbol{d}_{n}\right)$,
and
\[
\varsigma_{k,n,l}=3|u_{k,n,l}|^{2}(\sum_{k=1}^{K}p_{k}|\boldsymbol{h}_{n,k}^{H}\widetilde{\boldsymbol{v}}_{n,l}|^{2}+\left\Vert \widetilde{\boldsymbol{v}}_{n,l}\right\Vert ^{2}),\forall n,l.
\]
In the following, we use the Lagrange dual method to solve subproblem
(\ref{eq:subpd}). The Lagrange function for (\ref{eq:subpd}) is
\begin{align*}
\mathfrak{L}(\boldsymbol{d},\boldsymbol{\lambda}) & =\sum_{k=1}^{K}\sum_{n=1}^{N}\sum_{l=1}^{L}\mu_{k}w_{k}\varsigma_{k,n,l}4^{-d_{n,l}},\\
 & +\sum_{n=1}^{N}\lambda_{n}(2B_{W}\sum_{l=1}^{L}d_{n,l}-C_{n}),\forall\boldsymbol{d}\geq\boldsymbol{0},
\end{align*}
where $\boldsymbol{\lambda}=\left[\lambda_{1},...,\lambda_{N}\right]^{T}$
is the Lagrange multiplier vector. Since subproblem (\ref{eq:subpd})
is convex, the optimal quantization bits allocation can be obtained
by solving the KKT conditions as
\begin{equation}
d_{n,l}^{*}\left(\lambda_{n}\right)=\left[\frac{\log2B_{W}\lambda_{n}-\log(\log4\sum_{k=1}^{K}\mu_{k}w_{k}\varsigma_{k,n,l})}{\log4}\right]^{+},\label{eq:optd}
\end{equation}
$\forall n,l$, where the optimal Lagrange multiplier $\lambda_{n}\geq0$
is chosen such that $2B_{W}\sum_{l=1}^{L}d_{n,l}^{*}\left(\lambda_{n}\right)=C_{n}$.

\subsubsection{Convergence of the Short-term BC Algorithm}

The short-term BC algorithm is an instance of the MM algorithm in
\cite{Jacobson_TIP07_MM}. From Theorem 4.4 in \cite{Jacobson_TIP07_MM},
we have the following result.
\begin{thm}
[Convergence of Short-term BC Algorithm]\label{thm:Convergence-of-Short-term}Suppose
Problem (\ref{eq:WMMSE}) has a discrete set of stationary points.
As $J\rightarrow\infty$, the short-term BC algorithm converges to
a stationary point $\left(\boldsymbol{\beta}^{*},\boldsymbol{v}^{*},\boldsymbol{d}^{*},\boldsymbol{u}^{*},\boldsymbol{w}^{*}\right)$
of Problem (\ref{eq:WMMSE}). Moreover, 
\[
\left(\boldsymbol{x}-\boldsymbol{x}^{*}\right)^{T}\mathbf{J}_{\boldsymbol{x}}\left(\boldsymbol{\theta}^{*},\boldsymbol{x}^{*};\boldsymbol{H}\right)\boldsymbol{\mu}\leq0,\forall\boldsymbol{x}\in\mathcal{\widetilde{\mathcal{X}}},
\]
where $\boldsymbol{x}^{*}=\left[\boldsymbol{p}^{*T},\boldsymbol{v}^{*T},\boldsymbol{d}^{*T},\boldsymbol{u}^{*T}\right]^{T}$,
and $p_{k}^{*}=\left|\beta_{k}^{*}\right|^{2},\forall k$. 
\end{thm}

In Theorem \ref{thm:Convergence-of-Short-term}, we assume that Problem
(\ref{eq:WMMSE}) has a discrete set of stationary points to ensure
the convergence of to a single stationary point. Even if this condition
is violated, the short-term BC algorithm can still converge to an
invariant set of stationary points of (\ref{eq:WMMSE}) in the worst
case. However, such worst-case scenario rarely occurs in practice
\cite{Jacobson_TIP07_MM,Lee_CLT2016_GDlocal}. To ensure the exact
convergence of the overall Algorithm 1 to stationary solutions, we
need to let $J_{t}\rightarrow\infty$, as $t\rightarrow\infty$. As
$t\rightarrow\infty$, the output of the short-term BC algorithm $\boldsymbol{x}^{J_{\infty}}\left(\boldsymbol{\mu},\boldsymbol{\theta},\boldsymbol{H}\right)\triangleq\lim_{t\rightarrow\infty}\boldsymbol{x}^{J_{t}}\left(\boldsymbol{\mu},\boldsymbol{\theta},\boldsymbol{H}\right)$
is well defined only when it converges to a single stationary point.
Therefore, in the convergence analysis of Algorithm 1 in the next
subsection, we will assume that the short-term BC algorithm converges
to a single stationary point w.p.1. If we allow approximate convergence
by running the short-term BC algorithm for only a finite number of
iterations i.e., $\lim_{t\rightarrow\infty}J_{t}=J_{\infty}<\infty$
(which is always the case in practice), then $\boldsymbol{x}^{J_{\infty}}\left(\boldsymbol{\mu},\boldsymbol{\theta},\boldsymbol{H}\right)$
is always well defined and the assumption that Problem (\ref{eq:WMMSE})
has a discrete set of stationary points can be removed.

\subsection{Convergence Analysis}

In this section, we establish the local convergence of BC-SSCA to
stationary solutions. Due to the complex coupling between the short-term
and long-term control variables, the convergence of long-term control
variable depends heavily on the properties of the short-term solution
$\boldsymbol{x}^{J_{t}}\left(\boldsymbol{\mu}^{t},\boldsymbol{\theta}^{t},\boldsymbol{H}(i)\right)$
found by the short-term BC algorithm. Since there is no closed-form
characterization of $\boldsymbol{x}^{J_{t}}\left(\boldsymbol{\mu}^{t},\boldsymbol{\theta}^{t},\boldsymbol{H}(i)\right)$,
it is difficult to prove the convergence of BC-SSCA, which gives rise
to the following challenge. 
\begin{center}
\fbox{\begin{minipage}[t]{0.96\columnwidth}%
\begin{challenge}\label{chl:Deterministic-Restriction-2}For the
BC-SSCA which involves an iterative short-term BC algorithm without
closed-form characterization, it is non-trivial to establish its local
convergence to stationary solutions.\end{challenge}\vspace{-5bp}
\end{minipage}}
\par\end{center}

To address Challenge \ref{chl:Deterministic-Restriction-2}, we need
to make the following assumptions on the parameters $\left\{ \rho_{t},\gamma_{t},J_{t}\right\} $.

\begin{assumption}[Assumptions on $\left\{ \rho_{t},\gamma_{t},J_{t}\right\} $]\label{asm:convS}$\:$
\begin{enumerate}
\item $\rho_{t}\rightarrow0$, $\frac{1}{\rho_{t}}\leq O\left(t^{\kappa}\right)$
for some $\kappa\in\left(0,1\right)$, $\sum_{t}\left(\rho_{t}\right)^{2}<\infty$\footnote{We use $O\left(\cdot\right)$ to denote the Big O notation. Therefore,
$\frac{1}{\rho_{t}}\leq O\left(t^{\kappa}\right)$ means that $\limsup_{t\rightarrow\infty}\frac{t^{-\kappa}}{\rho_{t}}<\infty$.}.
\item $\gamma_{t}\rightarrow0$, $\sum_{t}\gamma_{t}=\infty$, $\sum_{t}\left(\gamma_{t}\right)^{2}<\infty$.
\item $\lim_{t\rightarrow\infty}\gamma_{t}/\rho_{t}=0$.
\item For any fixed $t>0$, $J_{t}<\infty$; and $J_{t}\rightarrow\infty$,
as $t\rightarrow\infty$.
\end{enumerate}
\end{assumption}

With Assumption \ref{asm:convS}, we first prove a key lemma which
establishes the convergence of the recursive approximations $\hat{\boldsymbol{r}}^{t},$
$\mathbf{f}^{t}$ and the surrogate function $\bar{f}^{t}\left(\boldsymbol{\theta}\right)$.
\begin{lem}
[Convergence of $\hat{\boldsymbol{r}}^{t}$, $\mathbf{f}^{t}$ and
$\bar{f}^{t}\left(\boldsymbol{\theta}\right)$]\label{lem:Convergence-of-Surro}Under
Assumption \ref{asm:convS}, we have
\begin{align}
\lim_{t\rightarrow\infty}\left|\hat{r}_{k}^{t}-\overline{r}_{k}\left(\boldsymbol{\theta}^{t},\Omega^{J_{t}}\left(\boldsymbol{\mu}^{t},\boldsymbol{\theta}^{t}\right)\right)\right| & =0,\label{eq:limtr}\\
\lim_{t\rightarrow\infty}\left\Vert \mathbf{f}^{t}-\nabla_{\boldsymbol{\theta}}g\left(\overline{\boldsymbol{r}}\left(\boldsymbol{\theta}^{t},\Omega^{J_{t}}\left(\boldsymbol{\mu}^{t},\boldsymbol{\theta}^{t}\right)\right)\right)\right\Vert  & =0,\label{eq:limtf}\\
\lim_{t\rightarrow\infty}\left\Vert \boldsymbol{\mu}^{t}-\nabla_{\overline{\boldsymbol{r}}}g\left(\overline{\boldsymbol{r}}\left(\boldsymbol{\theta}^{t},\Omega^{J_{t}}\left(\boldsymbol{\mu}^{t},\boldsymbol{\theta}^{t}\right)\right)\right)\right\Vert  & =0,\label{eq:limtmu}
\end{align}
where $\Omega^{J}\left(\boldsymbol{\mu},\boldsymbol{\theta}\right)=\left\{ \boldsymbol{x}^{J}\left(\boldsymbol{\mu},\boldsymbol{\theta},\boldsymbol{H}\right),\forall\boldsymbol{H}\right\} ,$
and $\boldsymbol{x}^{J}\left(\boldsymbol{\mu},\boldsymbol{\theta},\boldsymbol{H}\right)$
is the output of Algorithm 2 with input $J$, $\boldsymbol{\mu}$,
$\boldsymbol{\theta}$ and $\boldsymbol{H}$. Moreover, consider a
subsequence $\left\{ \boldsymbol{\mu}^{t_{j}},\boldsymbol{\theta}^{t_{j}}\right\} _{j=1}^{\infty}$
converging to a limiting point $\left(\boldsymbol{\mu}^{*},\boldsymbol{\theta}^{*}\right)$,
and define a function 
\begin{align*}
\hat{f}\left(\boldsymbol{\theta}\right) & \triangleq g\left(\overline{\boldsymbol{r}}\left(\boldsymbol{\theta}^{*},\Omega^{J_{\infty}}\left(\boldsymbol{\mu}^{*},\boldsymbol{\theta}^{*}\right)\right)\right)-\tau\left\Vert \boldsymbol{\theta}-\boldsymbol{\theta}^{*}\right\Vert ^{2}\\
 & +\nabla_{\boldsymbol{\theta}}^{T}g\left(\overline{\boldsymbol{r}}\left(\boldsymbol{\theta}^{*},\Omega^{J_{\infty}}\left(\boldsymbol{\mu}^{*},\boldsymbol{\theta}^{*}\right)\right)\right)\left(\boldsymbol{\theta}-\boldsymbol{\theta}^{*}\right),
\end{align*}
where $\Omega^{J_{\infty}}\left(\boldsymbol{\mu},\boldsymbol{\theta}\right)=\left\{ \boldsymbol{x}^{J_{\infty}}\left(\boldsymbol{\mu},\boldsymbol{\theta},\boldsymbol{H}\right),\forall\boldsymbol{H}\right\} ,$
and $\boldsymbol{x}^{J_{\infty}}\left(\boldsymbol{\mu},\boldsymbol{\theta},\boldsymbol{H}\right)$
is the stationary point of $\mathcal{P}_{S}\left(\boldsymbol{\mu},\boldsymbol{\theta},\boldsymbol{H}\right)$
found by Algorithm 2 (i.e., run Algorithm 2 until convergence to a
stationary point). Then, almost surely, we have
\begin{align}
\lim_{j\rightarrow\infty}\bar{f}^{t_{j}}\left(\boldsymbol{\theta}\right) & =\hat{f}\left(\boldsymbol{\theta}\right),\:\forall\boldsymbol{\theta}\in\Theta.\label{eq:ghfhead}
\end{align}

\end{lem}
Please refer to Appendix \ref{subsec:Proof-of-Lemma} for the proof.
The motivation for some key assumptions and the intuition behind Lemma
\ref{lem:Convergence-of-Surro} are explained below. From the recursive
update for $\hat{r}_{k}^{t}$ in (\ref{eq:fit}), $\hat{r}_{k}^{t}$
is roughly obtained by averaging the instantaneous rates over a time
window of size $\frac{1}{\rho_{t}}$. Since $\boldsymbol{\theta}^{t}$
is changing over time $t$, $\hat{r}_{k}^{t}$ may not converge to
$\overline{r}_{k}\left(\boldsymbol{\theta}^{t},\Omega^{J_{t}}\left(\boldsymbol{\mu}^{t},\boldsymbol{\theta}^{t}\right)\right)$
in general. However, if $\lim_{t\rightarrow\infty}\gamma_{t}/\rho_{t}=0$,
it follows from (\ref{eq:updatext}) that $\boldsymbol{\theta}^{t}$
is almost unchanged during the time window $\frac{1}{\rho_{t}}$ (i.e.,
$\boldsymbol{\theta}^{t-\frac{1}{\rho_{t}}}\approx\boldsymbol{\theta}^{t-\frac{1}{\rho_{t}}+1}\approx...\approx\boldsymbol{\theta}^{t-1}\approx\boldsymbol{\theta}^{t}$)
for sufficiently large $t$, and thus $\hat{r}_{k}^{t}$ will converge
to $\overline{r}_{k}\left(\boldsymbol{\theta}^{t},\Omega^{J_{t}}\left(\boldsymbol{\mu}^{t},\boldsymbol{\theta}^{t}\right)\right)$
as $t\rightarrow\infty$. The same assumption has also been made in
single-timescale stochastic optimization (with long-term control variable
only) \cite{Yang_TSP2016_SSCA} for the same reason. Another standard
technical assumption $\sum_{t}\rho_{t}=\infty$ is required in single-timescale
stochastic optimization. However, for the considered two-timescale
stochastic optimization in which the short-term control variables
$\boldsymbol{x}\left(i\right)$ are obtained by an iterative short-term
BC algorithm, a slightly stronger condition $\frac{1}{\rho_{t}}\leq O\left(t^{\kappa}\right)$
with $\kappa\in\left(0,1\right)$ than $\sum_{t}\rho_{t}=\infty$
is required. With Lemma \ref{lem:Convergence-of-Surro}, the following
convergence theorem can be proved.
\begin{thm}
[Convergence of Algorithm 1]\label{thm:Convergence-of-Algorithm1}Suppose
Assumption \ref{asm:convS} is satisfied. Let $\left\{ \boldsymbol{\mu}^{t_{j}},\boldsymbol{\theta}^{t_{j}}\right\} _{j=1}^{\infty}$
denote any subsequence of iterates generated by Algorithm 1 that converges
to a limiting point $\left(\boldsymbol{\mu}^{*},\boldsymbol{\theta}^{*}\right)$.
Then we almost surely have
\[
\boldsymbol{\mu}^{*}=\nabla_{\overline{\boldsymbol{r}}}g\left(\overline{\boldsymbol{r}}^{*}\right),
\]
\begin{equation}
\left(\boldsymbol{\theta}-\boldsymbol{\theta}^{*}\right)^{T}\nabla_{\boldsymbol{\theta}}g\left(\overline{\boldsymbol{r}}\left(\boldsymbol{\theta}^{*},\Omega^{J_{\infty}}\left(\boldsymbol{\mu}^{*},\boldsymbol{\theta}^{*}\right)\right)\right)\leq0,\label{eq:ooutconv}
\end{equation}
$\forall\boldsymbol{\theta}\in\Theta$, where $\overline{\boldsymbol{r}}^{*}=\overline{\boldsymbol{r}}\left(\boldsymbol{\theta}^{*},\Omega^{J_{\infty}}\left(\boldsymbol{\mu}^{*},\boldsymbol{\theta}^{*}\right)\right)$.
Moreover,
\begin{equation}
\lim_{j\rightarrow\infty}\left(\boldsymbol{x}-\boldsymbol{x}\left(i\right)\right)^{T}\mathbf{J}_{\boldsymbol{x}}\left(\boldsymbol{\theta}^{*},\boldsymbol{x}\left(i\right);\boldsymbol{H}(i)\right)\nabla_{\overline{\boldsymbol{r}}}g\left(\overline{\boldsymbol{r}}^{*}\right)=0,\label{eq:inerconv}
\end{equation}
where $\boldsymbol{x}\left(i\right)\triangleq\boldsymbol{x}^{J_{t_{j}}}\left(\boldsymbol{\mu}^{t_{j}},\boldsymbol{\theta}^{t_{j}},\boldsymbol{H}(i)\right),\forall i\in\left[t_{j}T_{s}+1,\left(t_{j}+1\right)T_{s}\right]$.
\end{thm}

Please refer to Appendix \ref{subsec:Proof-of-Theorem} for the proof.
According to Theorem \ref{thm:Convergence-of-Algorithm1}, as $j\rightarrow\infty$,
for any $i\in\left[t_{j}T_{s}+1,\left(t_{j}+1\right)T_{s}\right]$,
the short-term solution $\boldsymbol{x}\left(i\right)$ found by the
short-term BC algorithm satisfies the stationary condition in (\ref{eq:KKTs}).
Moreover, the limiting point $\boldsymbol{\theta}^{*}$ generated
by Algorithm 1 also satisfies the stationary condition in (\ref{eq:fix2}).
Therefore, Algorithm 1 converges to stationary solutions of the two-timescale
Problem $\mathcal{P}$.

\subsection{Computational Complexity\label{subsec:Computational-Complexity}}

In this subsection, we compare the computational complexity of the
proposed BC-SSCA algorithm with the following baseline schemes.
\begin{itemize}
\item \textbf{Baseline 1 - Spatial-compression-and-forward (SCF)} \cite{Liu_TSP2015_SCFCRAN}:
This is the SCF scheme in \cite{Liu_TSP2015_SCFCRAN} with the consideration
of the CSI delay. 
\item \textbf{Baseline 2 - Analog SCF} \textbf{(A-SCF)} \cite{Combi_EuCNC2017_FACRAN}:\textbf{
}This is the fully-analog spatial-compression-and-forward\textbf{
}scheme in \cite{Combi_EuCNC2017_FACRAN}. 
\item \textbf{Baseline 3 - Slow-timescale SCF} \textbf{(S-SCF)}:\textbf{
}This scheme is obtained by removing the short-term optimization in
the proposed scheme. 
\end{itemize}
We first analyze the complexity of the proposed BC-SSCA algorithm.
The complexity order for other schemes can be obtained similarly. 

\textbf{Complexity order of the short-term BC algorithm:} In each
iteration of the short-term BC algorithm, we solve the subproblems
for the five blocks of variables in three steps:

1) In Step 1, the calculation of $\left(\sum_{l=1}^{K}\widetilde{\boldsymbol{V}}^{H}\boldsymbol{h}_{l}\left|\beta_{l}\right|^{2}\boldsymbol{h}_{l}^{H}\widetilde{\boldsymbol{V}}+\widetilde{\boldsymbol{V}}^{H}\widetilde{\boldsymbol{V}}+\boldsymbol{Q}\right)^{-1}\in\mathbb{C}^{NL\times NL}$
and $\boldsymbol{u}_{k}$ needs $O(N^{3}L^{3})$ and $O(N^{2}L^{3})$
floating point operations (FPOs), respectively. Moreover, we only
need $O(NL)$ FPOs to compute $w_{k}=\left(1-\boldsymbol{u}_{k}^{H}\widetilde{\boldsymbol{V}}^{H}\boldsymbol{h}_{k}\beta_{k}\right)^{-1}$,
since $\boldsymbol{u}_{k}$ and $\widetilde{\boldsymbol{V}}^{H}\boldsymbol{h}_{k}\beta_{k}$
have been already been calculated previously. Similarly, the calculation
of $\beta_{k},\forall k$ in (\ref{eq:upbeta}) needs $O(1)$ FPOs.

2) In Step 2, the computation complexity of updating $\bm{v}$ is
dominated by the inversion of $\bm{B}$ and is given by $O(N^{3}L^{3}S^{3})$.

3) In Step 3, the bisection method to find the Lagrangian parameter
$\lambda_{n}$ requires $O(1)$ iterations to achieve certain accuracy
and $L$ multiplications are performed in each iteration. Therefore,
the computation complexity of updating $\bm{d}$ is $O(S+L).$

Based on the above analysis, the computation complexity of the short-term
BC algorithm is:
\[
C_{S}\triangleq O\left(N^{3}L^{3}+N^{3}L^{3}S^{3}+L\right).
\]

\textbf{Complexity order of the long-term control optimization:} The
computation complexity is dominated by the updating $\mathbf{f}$
and computing the $\mathbf{J}_{\boldsymbol{\theta}}\left(\boldsymbol{\theta},\boldsymbol{x};\boldsymbol{H}\right)$,
whose complexity order is $C_{L}=O(N^{2}ML+MSNK)$. 

\textbf{Overall complexity order of BC-SSCA:} Since each frame consists
of $T_{s}$ time slots, the overall complexity order of the proposed
BC-SSCA algorithm is $C_{BC-SSCA}\triangleq O\left(\text{\ensuremath{\frac{C_{L}}{T_{s}}}}+C_{S}\right)$. 

\textbf{Comparison of complexity orders: }In Table \ref{tab:Asymptotic-complexity},
we summarize the complexity orders of different schemes. As seen from
Table \ref{tab:Asymptotic-complexity}, since $M\gg S\geq L$, the
proposed THCF scheme has much lower complexity than that of both the
SCF scheme and the A-SCF scheme. Although the S-SCF scheme provides
a lower computation complexity than that of the proposed THCF scheme,
the performance is in general much worse. Consequently, our proposed
THCF scheme offers a better trade-off between complexity and performance. 

\begin{table}
\begin{onehalfspace}
\begin{centering}
\begin{tabular}{|c|c|}
\hline 
Schemes & Complexity order\tabularnewline
\hline 
\hline 
THCF scheme & $O(\text{\ensuremath{\frac{N^{2}ML+MSNK}{T_{s}}}+}N^{3}L^{3}S^{3})$\tabularnewline
\hline 
SCF scheme & $O(N^{3}L^{3}M^{3})$\tabularnewline
\hline 
A-SCF scheme & $O(N^{3}L^{3}M^{3})$\tabularnewline
\hline 
S-SCF scheme & $O(\ensuremath{\frac{N^{2}ML+MSNK}{T_{s}}})$\tabularnewline
\hline 
\end{tabular}
\par\end{centering}
\end{onehalfspace}
\caption{\label{tab:Asymptotic-complexity}Complexity orders for different
schemes.}
\end{table}

\subsection{Implementation Consideration}

At the beginning of each frame, the BBU needs to send $\boldsymbol{\theta}_{n}$
to RRH $n$. Moreover, at each time slot, the BBU needs to send $\boldsymbol{V}_{n},\boldsymbol{d}_{n}$
to RRH $n$ and $p_{k}$ to user $k$. In practice, each of these
control variables $\boldsymbol{\theta}_{n},\boldsymbol{V}_{n},p_{k}$
needs to be quantized using, e.g., a codebook based method, before
sending them to the RRHs or users (note that $\boldsymbol{d}_{n}$
is already an integer). The detailed codebook design for each control
variable is out of the scope of this paper. In the simulations, we
observe that the loss due to the quantization of $\boldsymbol{\theta}_{n}$
is already small under the simple uniform scalar quantization with
only 3-bits quantization for each element. Since $\boldsymbol{\theta}_{n}$
is adaptive to the channel statistics and is only updated once per
frame ($T_{f}$ time slots), the signaling overhead for communicating
the quantized $\boldsymbol{\theta}_{n}$ is relatively small. On the
other hand, by adapting the short-term control variables $\boldsymbol{V}_{n},\boldsymbol{d}_{n}$
and $p_{k}$ to the effective channel, these variables are updated
once per time slot. Note that $\boldsymbol{V}_{n},\boldsymbol{d}_{n}$
can be conveyed to RRH $n$ via a dedicated high-speed fronthaul link
and this may not cause too much overhead compared to the amount of
data symbols that needs to be send to RRH $n$ (since each time slot
may contain a large number of data symbols). Although $p_{k}$ needs
to be send to each user $k$ via the downlink wireless channel with
a lower capacity compared to the fronthaul link, it is only a scalar
and thus the resulting signaling overhead is still acceptable in practice. 

For users with higher mobility, the channel coherence time is smaller
and the signaling overhead for sending $\boldsymbol{V}_{n},\boldsymbol{d}_{n}$
to RRH $n$ may become unacceptable. In this case, we can simply use
a distributed digital filter $\boldsymbol{V}_{n}$ as in \cite{Liu_TSP2015_SCFCRAN}
such that each RRH can independently determine its digital filter
$\boldsymbol{V}_{n}$ based on the covariance matrix of its received
pilot signal that can be obtained locally at each RRH by sending uplink
pilots from the users. Please refer to \cite{Liu_TSP2015_SCFCRAN}
for the detailed design of the distributed digital filter $\boldsymbol{V}_{n}$.
In addition, we can simply use a uniform quantization (i.e., $d_{n,l}=\frac{C_{n}}{2BL},\forall n,l$)
to avoid the signaling overhead of sending $\boldsymbol{d}_{n}$ to
RRH $n$. Therefore, the proposed algorithm framework can be easily
modified to achieve a good tradeoff between the performance and signaling
overhead for practical implementations.
\begin{rem}
In this paper, we focus on fast power control and thus the power allocation
vector $\boldsymbol{p}$ is a short-term control variable. In practice,
for high mobility user with smaller channel coherence time, we may
switch to slow power control to avoid frequent fast power control
signaling. The proposed algorithm can be easily modified to consider
slow power control by removing the power allocation from the short-term
control optimization and modifying the surrogate function for long-term
control optimization to include the power allocation as a long-term
variable.
\end{rem}

\section{Simulation Results and Discussions\label{sec:Simulation-Results}}

Consider a C-RAN with 4 RRHs placed in a circle cell of radius 500
m. There are 8 users randomly distributed in the cell. The channel
bandwidth is 1 MHz. As in \cite{Park_TSP17_THP}, we adopt a geometry-based
channel model with a \textit{half-wavelength space} ULA for simulations.
The channel vector between RRH $n$ and user $k$ can be expressed
as $\boldsymbol{h}_{n,k}=\sum_{i=1}^{N_{p}}\alpha_{n,k,i}\boldsymbol{\textrm{a}}\left(\varphi_{n,k,i}\right)$,
where $\boldsymbol{\textrm{a}}\left(\varphi\right)$ is the array
response vector, $\varphi_{n,k,i}$'s are Laplacian distributed with
an angle spread $\sigma_{\textrm{AS}}=10$, $\alpha_{n,k,i}\sim\mathcal{CN}\left(0,\sigma_{n,k,i}^{2}\right)$,
$\sigma_{n,k,i}^{2}$ are randomly generated from an exponential distribution
and normalized such that $\sum_{i=1}^{N_{p}}\sigma_{n,k,i}^{2}=G_{n,k}$,
$G_{n,k}$ is the average channel gain determined by the pathloss
model $30.6+36.7\log10\left(\textrm{dist}_{n,k}\right)$ \cite{3gpp_Rel9},
and $\textrm{dist}_{n,k}$ is the distance between RRH $n$ and user
$k$ in meters. Unless otherwise specified, we consider $M=64$ antennas,
$S=16$ RF chains and $N_{p}=6$ channel paths for each RRH. The power
spectral density of the background noise is -169 dBm/Hz. The transmit
power constraint for each user is $P_{k}=23$ dBm. There are $T_{s}=10$
time slots in each frame and the slot size is 1 ms. The coherence
time for the channel statistics is assumed to be 10 s \cite{Ingo_COST02_Svar}
. As in \cite{Liu_TSP14_RFprecoding,Liu_TSP2016_CSImassive}, we assume
that the CSI delay is proportional to the dimension of the channel
vector that is required at the BS, i.e., if the \textit{full-CSI delay}
(which is defined as the delay required to obtain the full channel
sample $\boldsymbol{H}_{n},\forall n$) is $\tau$ ms, then the \textit{effective-CSI
delay} (which is defined as the delay required to obtain the effective
channel $\boldsymbol{F}_{n}^{H}\boldsymbol{H}_{n},\forall n$) is
$\frac{S}{M}\tau$ ms. The carrier frequency is 2.14 GHz and the velocity
of users is 3 Km/h. The CSI delay is set to be $\tau=4$ ms except
for Fig. \ref{fig:CSI}.

We use PFS utility as an example to illustrate the advantage of the
proposed scheme. Three baseline schemes described in Section \ref{subsec:Computational-Complexity}
are considered for comparison. In the simulations, the performance
of the ideal SCF without CSI delay is also provided as a performance
upper bound.

\subsection{Convergence of the online BC-SSCA algorithm}

\begin{figure}
\begin{centering}
\includegraphics[width=0.95\columnwidth]{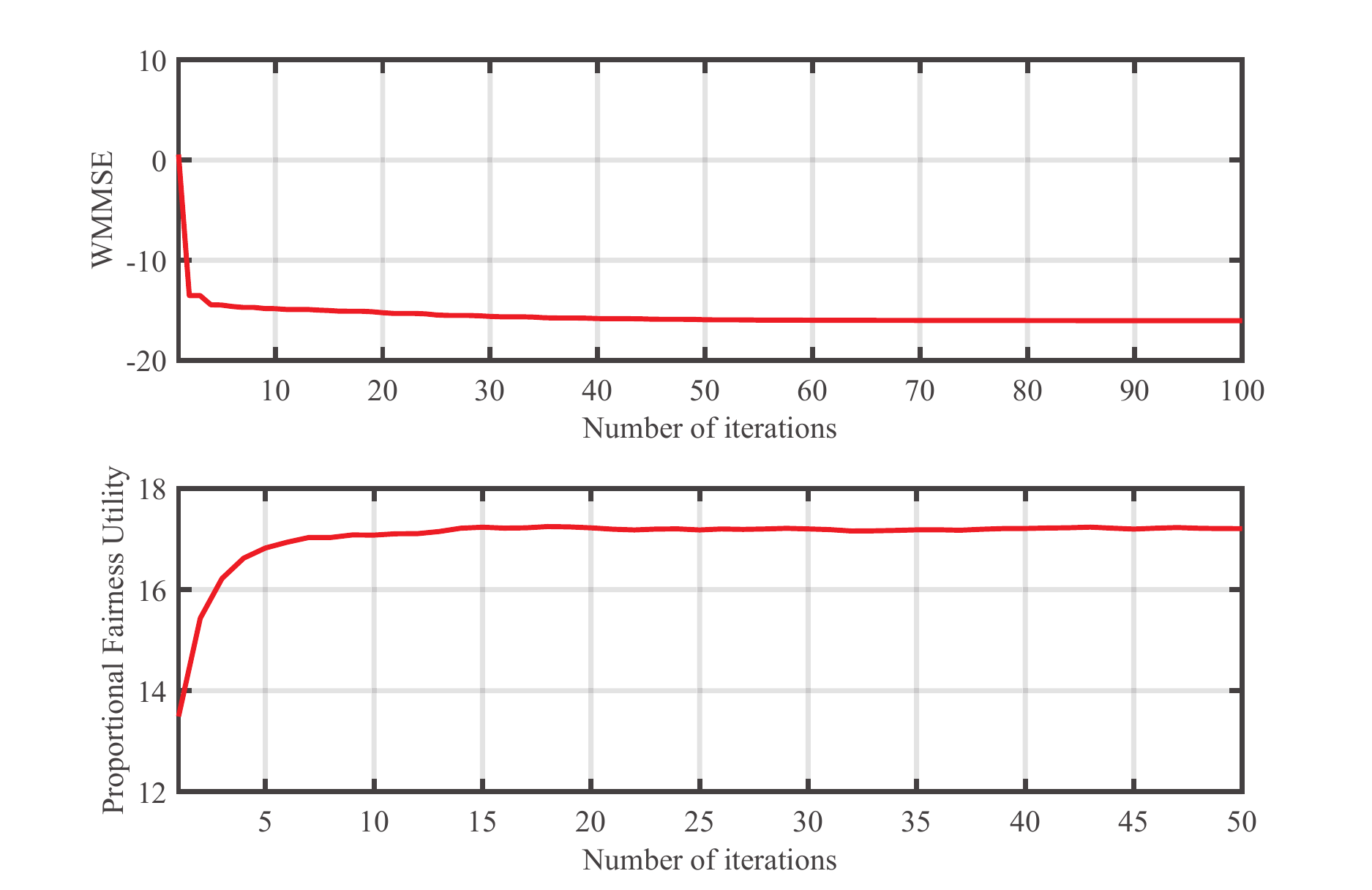}
\par\end{centering}
\caption{\label{fig:long}Convergence of the BC-SSCA algorithm. }
\end{figure}

In the upper subplot of Fig. \ref{fig:long}, we plot the objective
function of the short-term BC algorithm versus the iteration number.
The short-term BC algorithm converges within a few iterations. The
lower subplot illustrates the convergence behavior for the overall
BC-SSCA algorithm. It can be seen that BC-SSCA quickly converges to
a stationary solution.

\subsection{Performance versus the Fronthaul Link Capacity}

\begin{figure}
\begin{centering}
\includegraphics[width=0.95\columnwidth]{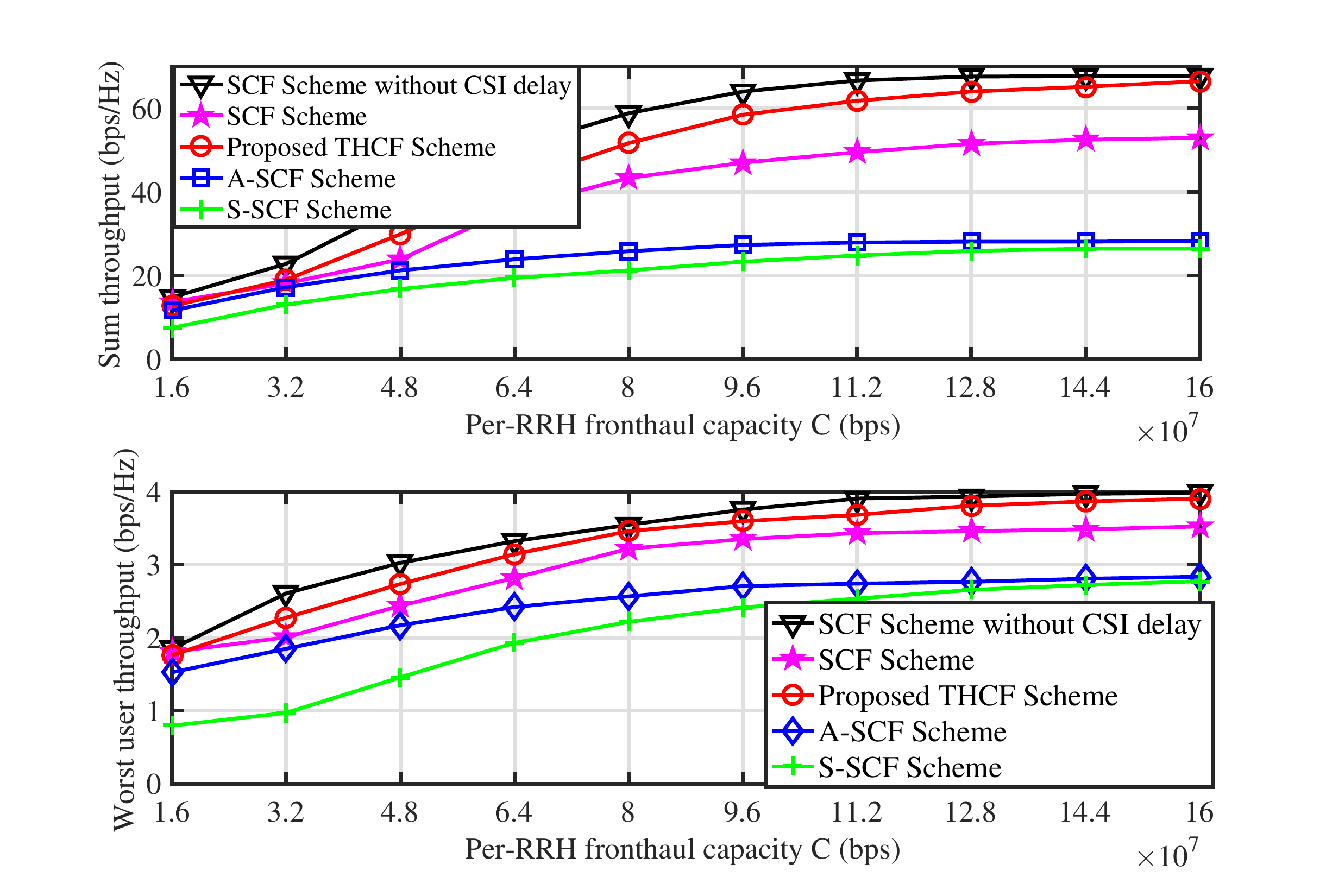}
\par\end{centering}
\caption{\label{fig:FC}Rate performance versus the per-RRH fronthaul capacity
$C$. }
\end{figure}

Fig. \ref{fig:FC} shows the performance comparison of different schemes
versus per-RRH fronthaul capacity $C$ varies from $C=16$ Mbps to
$C=160$ Mbps. It can be observed that the best performance of both
the sum throughput and worst user throughput are achieved by the SCF
scheme without CSI delay, followed by the proposed THCF scheme. Furthermore,
the proposed THCF scheme achieves significant gain over A-SCF and
S-SCF, which demonstrates the importance of hybrid analog-and-digital
processing and two-timescale joint optimization. When the fronthaul
capacity increases, the performance gap between the proposed THCF
scheme and the performance upper bound (SCF without CSI delay) becomes
smaller. Finally, it is observed that the performance of SCF is inferior
to the proposed THCF since the full-CSI delay is larger than the effective-CSI
delay.

\subsection{Performance versus the Number of Antennas per RRH}

\begin{figure}
\begin{centering}
\includegraphics[width=0.95\columnwidth]{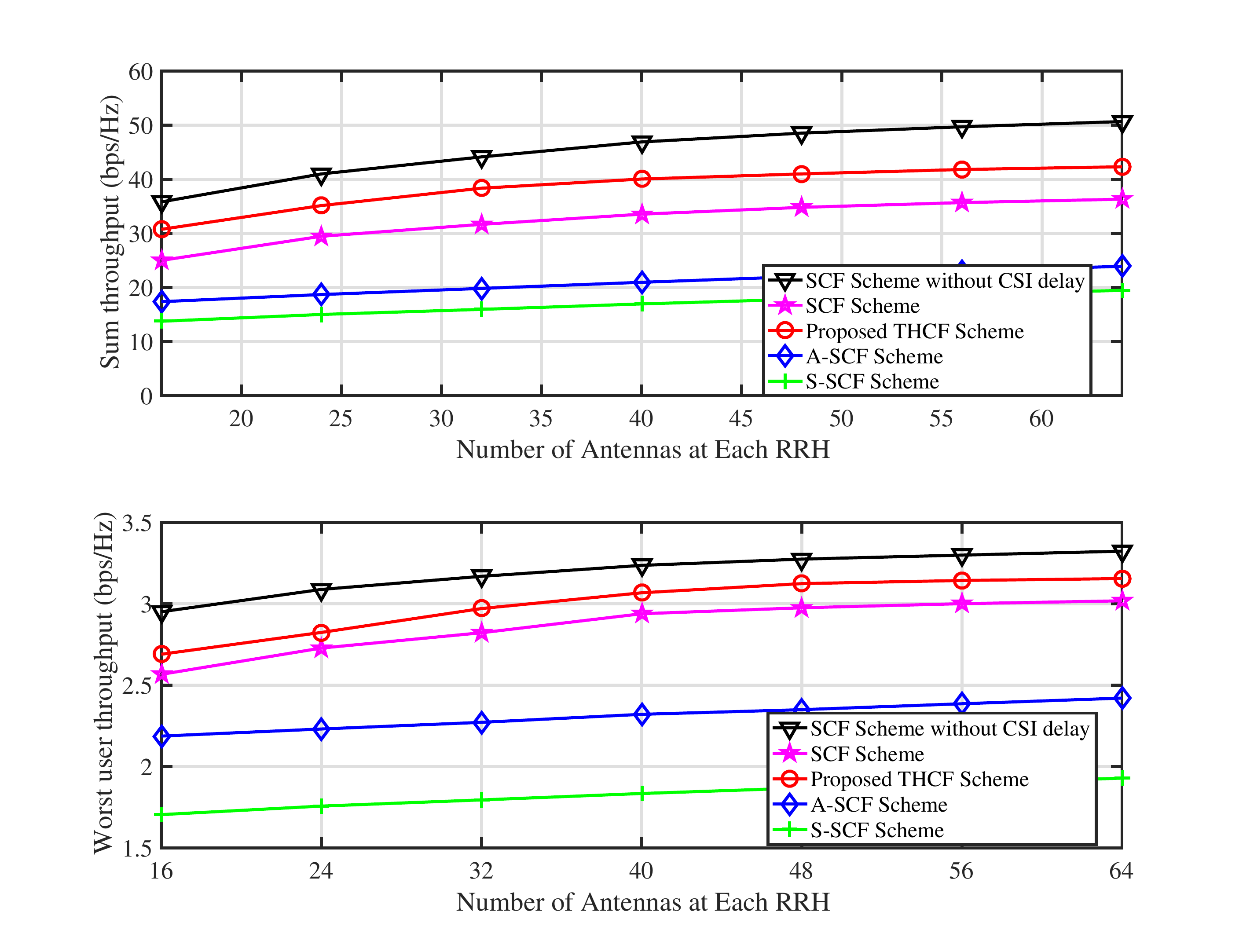}
\par\end{centering}
\caption{\label{fig:NR}Rate performance versus the per-RRH number of antennas
$M$.}
\end{figure}

In Fig. \ref{fig:NR} , we plot the rate performance versus the number
of antennas $M$ per-RRH, where the per-RRH fronthaul capacity is
fixed as $C=64$Mbps. We observe that the proposed THCF scheme achieves
a near-optimal performance when compared to the SCF scheme without
CSI delay (a performance upper bound) and outperforms the competing
schemes. Moreover, as $M$ increases, the performance gap between
the competing schemes becomes larger. Again, the SCF scheme without
CSI delay achieves the best PFS performance, but its hardware complexity
and implementation cost are much larger than the proposed THCF especially
when the number of antennas $M$ per-RRH is large. Moreover, when
there is CSI delay, the proposed THCF scheme will outperform the SCF
scheme. In contrast, with the proposed THCF scheme, we can enjoy the
huge array gain provided by the massive MIMO almost for free (i.e.,
the complexity and power consumption are similar to the C-RAN with
small-scale multi-antenna RRHs). This indicates that the proposed
THCF scheme achieves better tradeoff performance than other baselines.

\subsection{Performance versus the CSI Delay}

\begin{figure}
\begin{centering}
\includegraphics[width=0.95\columnwidth]{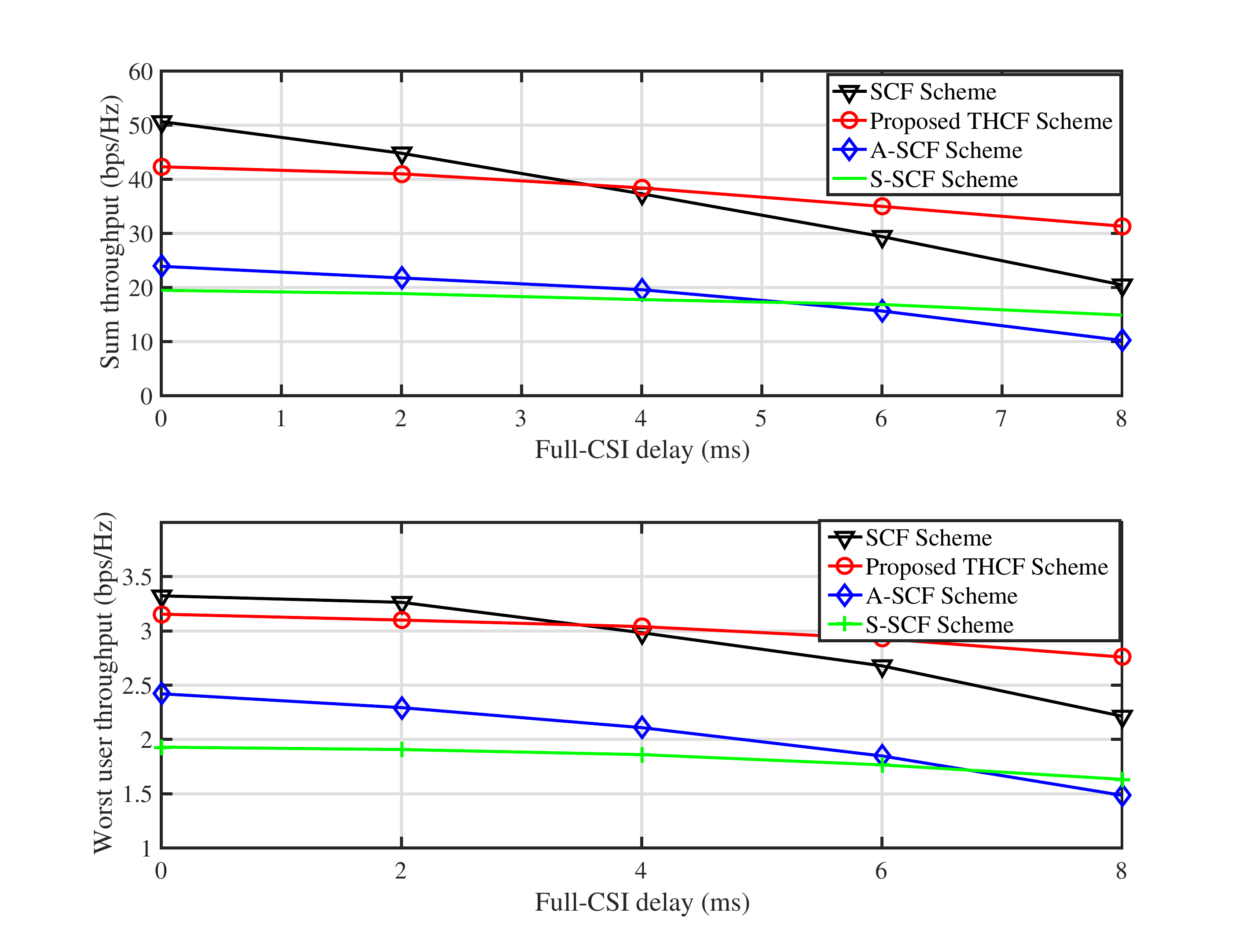}
\par\end{centering}
\caption{\label{fig:CSI}Rate performance versus the CSI delay. }
\end{figure}

In Fig. \ref{fig:CSI}, we plot the rate performance versus the CSI
delay, where the per-RRH fronthaul capacity is fixed as $C=64$Mbps.
We can see that as the CSI delay increases, the PFS of all schemes
decreases gradually. It is observed that the PFS achieved with the
proposed THCF scheme is higher than that achieved by the other schemes
for moderate and large full-CSI delay. This is because the performance
of the proposed THCF scheme is insensitive to the full-CSI delay.
Although the performance of the S-SCF scheme is also insensitive to
the full-CSI delay, its performance is still much worse than the proposed
THCF scheme due to the lack of optimal power control and quantization
bits allocation.

\subsection{Performance under Practical Implementation Consideration}

\begin{figure}
\begin{centering}
\includegraphics[width=0.95\columnwidth]{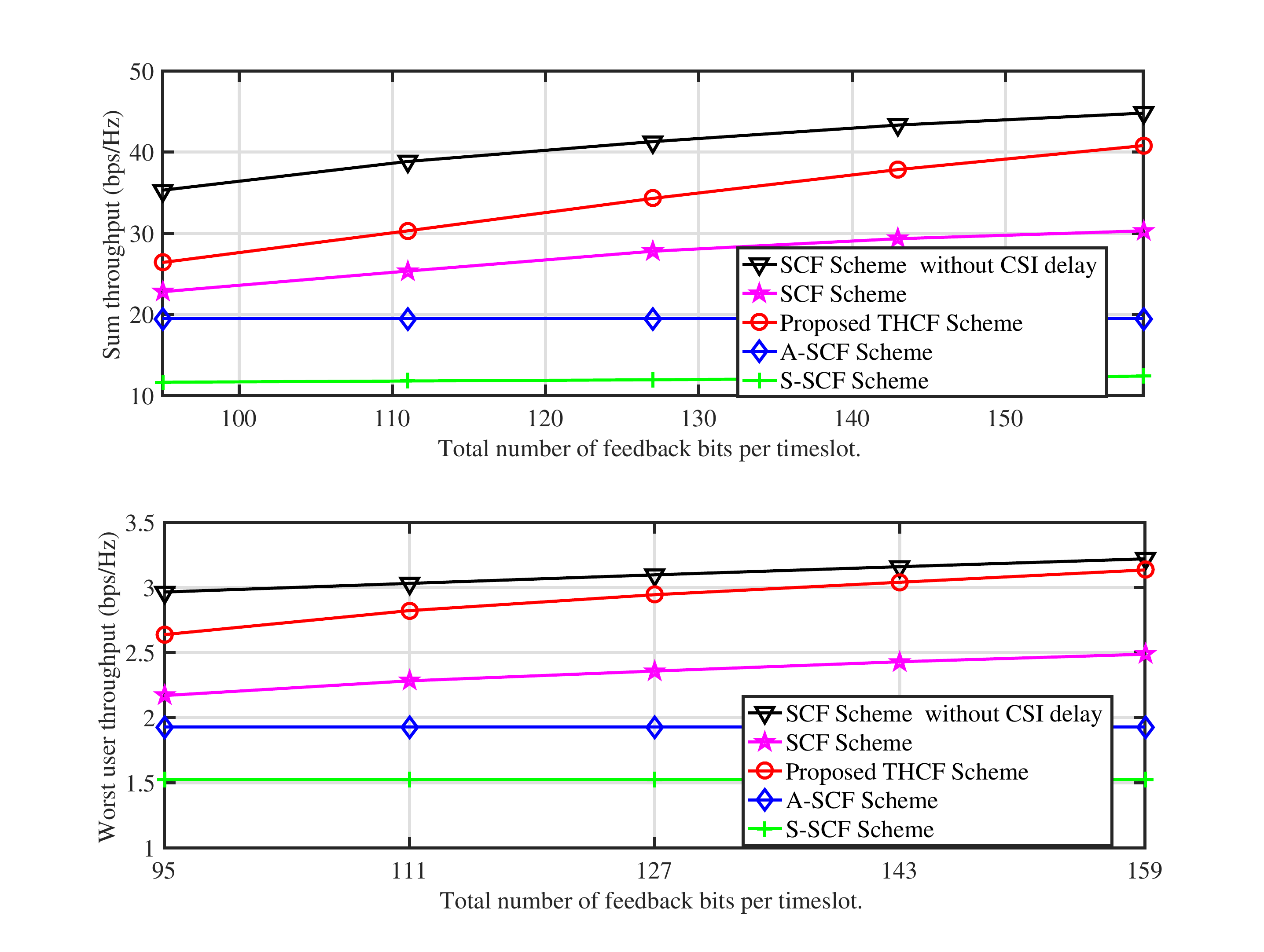}
\par\end{centering}
\caption{\label{fig:quantization}Rate performance versus the total feedback
bits per timeslot.}
\end{figure}

To illustrate the impact of quantization on the analog and digital
filtering matrices in practice, the performance of all schemes are
evaluated in Fig. \ref{fig:quantization} with different total numbers
of quantization bits. The per-RRH fronthaul capacity is fixed as $C=64$
Mbps. For THCF, we assume that $\boldsymbol{\theta}_{n}$ is quantized
using the simple uniform scalar quantization with only 3-bits quantization
for each element. Moreover, the digital filter $\boldsymbol{V}_{n}$
are quantized using a random vector quantization (RVQ) codebook. For
fair comparison, the total number of feedback bits per timeslot of
all schemes are set to be equal. As shown in Fig. \ref{fig:quantization},
the PFS utility of SCF and THCF increases with the total feedback
bits per timeslot. On the other hand, the PFS utility of S-SCF and
A-SCF almost remain constant when total feedback bits per timeslot
ranges from 95 to 159 bits. This is because for S-SCF, a total number
of 95 quantization bits corresponds to 3-bits quantization for each
phase in $\boldsymbol{\theta}_{n}$, since the analog filter is updated
at the slow timescale in S-SCF. In this case, the quantization loss
is already small compared to the unquantized case. For A-SCF, a total
number of 159 quantization bits only corresponds to less than 1-bit
quantization for each phase in $\boldsymbol{\theta}_{n}$, since the
analog filter is updated at the fast timescale in A-SCF. Therefore,
we can only use 1-bit quantization for each phase in A-SCF for the
entire range of feedback bits in the simulations. This also demonstrates
that the slow-timescale/two-timescale design can significantly reduce
the feedback overhead.

\section{Conclusion\label{sec:Conlusion}}

We propose a two-timescale hybrid compression and forward (THCF) scheme
to reduce the fronthaul consumption in Massive MIMO aided C-RAN. We
formulate the optimization of THCF as a general utility maximization
problem, and propose a BC-SSCA algorithm to find stationary solutions
of this two-stage non-convex stochastic optimization problem. At each
iteration, BC-SSCA first runs an iterative BC algorithm to find a
stationary point (up to a certain accuracy) of the short-term weighted
sum rate maximization subproblem associated with the observed channel
state. Then it updates the surrogate function for the objective of
the long-term analog spatial filtering problem based on the observed
channel state, the current iterate and the stationary point of the
short-term subproblem. Finally, it updates the long-term analog spatial
filter by solving the resulting convex approximation problem with
closed-form solution. We show that the BC-SSCA algorithm converges
to stationary solutions of the joint optimization problem almost surely.
Finally, simulations verify that the proposed BC-SSCA algorithm achieves
significant gain over existing solutions.

\appendix

\subsection{Jacobian Matrix of Instantaneous Rate\label{subsec:Jacobian-Matrix-of} }

For given channel state $\boldsymbol{H},$ the Jacobian matrix of
the instantaneous rate vector $r(\boldsymbol{\theta},\boldsymbol{x};\boldsymbol{H})$
with respect to $\boldsymbol{\theta}$ is
\[
\mathbf{J}_{r}\left(\boldsymbol{\theta},\boldsymbol{x};\boldsymbol{H}\right)=\left[\begin{array}{cccc}
\nabla_{\boldsymbol{\theta}}r_{1} & \nabla_{\boldsymbol{\theta}}r_{2} & \cdots & \nabla_{\boldsymbol{\theta}}r_{K}\end{array}\right],
\]
where $\nabla_{\boldsymbol{\theta}}r_{k}=[\nabla_{\boldsymbol{\theta}_{1}}^{T}r_{k},...,\nabla_{\boldsymbol{\theta}_{N}}^{T}r_{k}]^{T}$.
According to the matrix calculus and the chain rule, we can get
\begin{align*}
\nabla_{\boldsymbol{\theta}_{n}}r_{k} & =\frac{\boldsymbol{a}_{n,k}}{\Gamma_{k}}-\frac{\boldsymbol{a}_{n,-k}}{\Gamma_{-k}},
\end{align*}
where
\begin{multline*}
\Gamma_{k}=\sum_{l=1}^{K}p_{l}|\boldsymbol{u}_{k}^{H}\widetilde{\boldsymbol{V}}^{H}\boldsymbol{h}_{l}|^{2}+\|\boldsymbol{u}_{k}^{H}\widetilde{\boldsymbol{V}}^{H}\|^{2}+\boldsymbol{u}_{k}^{H}\boldsymbol{Q}\left(\boldsymbol{\theta},\boldsymbol{p},\boldsymbol{v},\boldsymbol{d}\right)\boldsymbol{u}_{k},\\
\Gamma_{-k}=\sum_{l\neq k}^{K}p_{l}|\boldsymbol{u}_{k}^{H}\widetilde{\boldsymbol{V}}^{H}\boldsymbol{h}_{l}|^{2}+\|\boldsymbol{u}_{k}^{H}\widetilde{\boldsymbol{V}}^{H}\|^{2}+\boldsymbol{u}_{k}^{H}\boldsymbol{Q}\left(\boldsymbol{\theta},\boldsymbol{p},\boldsymbol{v},\boldsymbol{d}\right)\boldsymbol{u}_{k},\\
\boldsymbol{a}_{n,k}=-\mathrm{Vec}(\mathfrak{R}[\sqrt{-1}\boldsymbol{F}_{n}^{*}\circ\boldsymbol{A}_{n,k}]),\\
\boldsymbol{a}_{n,-k}=-\mathrm{Vec}(\mathfrak{R}[\sqrt{-1}\boldsymbol{F}_{n}^{*}\circ\boldsymbol{A}_{n,-k}]),\\
\boldsymbol{A}_{n,k}=\sum_{l=1}^{K}p_{l}(\boldsymbol{h}_{n,l}\boldsymbol{u}_{n,k}^{H}\boldsymbol{V}_{n}^{H})\cdot(\boldsymbol{h}_{l}^{H}\widetilde{\boldsymbol{V}}\boldsymbol{u}_{k})+\boldsymbol{F}_{n}\boldsymbol{V}_{n}\boldsymbol{u}_{n,k}\boldsymbol{u}_{n,k}^{H}\boldsymbol{V}_{n}^{H}\\
+\sum_{l=1}^{L}\frac{3u_{n,k,l}^{H}u_{n,k,l}}{4^{d_{n,l}}}(\sum_{k}p_{k}\boldsymbol{h}_{n,k}\boldsymbol{h}_{n,k}^{H}\boldsymbol{F}_{n}\boldsymbol{v}_{n.l}\boldsymbol{v}_{n,l}^{H}+\boldsymbol{F}_{n}\boldsymbol{v}_{n,l}\boldsymbol{v}_{n,l}^{H}),\\
\boldsymbol{A}_{n,-k}=\sum_{l\neq k}^{K}p_{l}(\boldsymbol{h}_{n,l}\boldsymbol{u}_{n,k}^{H}\boldsymbol{V}_{n}^{H})\cdot(\boldsymbol{h}_{l}^{H}\widetilde{\boldsymbol{V}}\boldsymbol{u}_{k})+\boldsymbol{F}_{n}\boldsymbol{V}_{n}\boldsymbol{u}_{n,k}\boldsymbol{u}_{n,k}^{H}\boldsymbol{V}_{n}^{H}\\
+\sum_{l=1}^{L}\frac{3u_{n,k,l}^{H}u_{n,k,l}}{4^{d_{n,l}}}(\sum_{k}p_{k}\boldsymbol{h}_{n,k}\boldsymbol{h}_{n,k}^{H}\boldsymbol{F}_{n}\boldsymbol{v}_{n,l}\boldsymbol{v}_{n,l}^{H}+\boldsymbol{F}_{n}\boldsymbol{v}_{n,l}\boldsymbol{v}_{n,l}^{H}).
\end{multline*}

\subsection{Proof of Lemma \ref{lem:Convergence-of-Surro}\label{subsec:Proof-of-Lemma}}

The proof relies on the following lemma which characterizes the Lipschitz
continuity of $\boldsymbol{x}^{J}\left(\boldsymbol{\mu},\boldsymbol{\theta},\boldsymbol{H}\right)$
and $\overline{r}_{k}\left(\boldsymbol{\theta},\Omega^{J}\left(\boldsymbol{\mu},\boldsymbol{\theta}\right)\right),\forall k$
with respect to $\boldsymbol{\mu},\boldsymbol{\theta}$. 
\begin{lem}
\label{lem:Lipschitz-continuity-ofyf}Let $\boldsymbol{x}^{J}\left(\boldsymbol{\mu},\boldsymbol{\theta},\boldsymbol{H}\right)$
denote the output of Algorithm 2 with input $J$, $\boldsymbol{\mu}$,
$\boldsymbol{\theta}$ and $\boldsymbol{H}$. We have
\begin{align}
 & \left\Vert \boldsymbol{x}^{J}\left(\boldsymbol{\mu}_{1},\boldsymbol{\theta}_{1},\boldsymbol{H}\right)-\boldsymbol{x}^{J}\left(\boldsymbol{\mu}_{2},\boldsymbol{\theta}_{2},\boldsymbol{H}\right)\right\Vert \nonumber \\
\leq & B_{x}^{J}\sqrt{\left\Vert \boldsymbol{\mu}_{1}-\boldsymbol{\mu}_{2}\right\Vert ^{2}+\left\Vert \boldsymbol{\theta}_{1}-\boldsymbol{\theta}_{2}\right\Vert ^{2}},\label{eq:xcont}
\end{align}
\begin{align}
 & \left\Vert \overline{r}_{k}\left(\boldsymbol{\theta}_{1},\Omega^{J}\left(\boldsymbol{\mu}_{1},\boldsymbol{\theta}_{1}\right)\right)-\overline{r}_{k}\left(\boldsymbol{\theta}_{2},\Omega^{J}\left(\boldsymbol{\mu}_{2},\boldsymbol{\theta}_{2}\right)\right)\right\Vert \nonumber \\
\leq & B^{J}\sqrt{\left\Vert \boldsymbol{\mu}_{1}-\boldsymbol{\mu}_{2}\right\Vert ^{2}+\left\Vert \boldsymbol{\theta}_{1}-\boldsymbol{\theta}_{2}\right\Vert ^{2}},\label{eq:rcont}
\end{align}
for any $\boldsymbol{\mu}_{1}\boldsymbol{\mu}_{2}>0,\:\boldsymbol{\theta}_{1},\boldsymbol{\theta}_{2}\in\Theta$,
any inner iteration number $J\geq1$ and some constant $B_{x}>0$,
$B>0$.
\end{lem}
\begin{IEEEproof}
In Algorithm 2, the initial point $\boldsymbol{x}^{0}\left(\boldsymbol{\mu}_{1},\boldsymbol{\theta}_{1},\boldsymbol{H}\right)-\boldsymbol{x}^{0}\left(\boldsymbol{\mu}_{2},\boldsymbol{\theta}_{2},\boldsymbol{H}\right)$
is Lipschitz continuous with respect to $\boldsymbol{\mu}$, $\boldsymbol{\theta}$.
Moreover, each subproblem with respect to one short-term variable
has a closed-form solution which is also Lipschitz continuous with
respect to $\boldsymbol{\mu}$, $\boldsymbol{\theta}$ and the other
short-term variables. Therefore, after the first iterations, we have
\begin{align*}
 & \left\Vert \boldsymbol{x}^{1}\left(\boldsymbol{\mu}_{1},\boldsymbol{\theta}_{1},\boldsymbol{H}\right)-\boldsymbol{x}^{1}\left(\boldsymbol{\mu}_{2},\boldsymbol{\theta}_{2},\boldsymbol{H}\right)\right\Vert \\
\leq & B_{1}\sqrt{\left\Vert \boldsymbol{\mu}_{1}-\boldsymbol{\mu}_{2}\right\Vert ^{2}+\left\Vert \boldsymbol{\theta}_{1}-\boldsymbol{\theta}_{2}\right\Vert ^{2}},
\end{align*}
for some $B_{1}>0$, and after $J$ iteration, we have 
\begin{align*}
 & \left\Vert \boldsymbol{x}^{J}\left(\boldsymbol{\mu}_{1},\boldsymbol{\theta}_{1},\boldsymbol{H}\right)-\boldsymbol{x}^{J}\left(\boldsymbol{\mu}_{2},\boldsymbol{\theta}_{2},\boldsymbol{H}\right)\right\Vert \\
\leq & B_{1}B_{2}...B_{J}\sqrt{\left\Vert \boldsymbol{\mu}_{1}-\boldsymbol{\mu}_{2}\right\Vert ^{2}+\left\Vert \boldsymbol{\theta}_{1}-\boldsymbol{\theta}_{2}\right\Vert ^{2}}.
\end{align*}
for some $B_{j}>0,j=1,...,J$. Letting $B_{x}=\max_{j}B_{j}$, (\ref{eq:xcont})
is proved. Finally, (\ref{eq:rcont}) follows immediately from the
fact that $r_{k}\left(\boldsymbol{\theta},\boldsymbol{x},\boldsymbol{H}\right)$
is Lipschitz continuous with respect to $\boldsymbol{\theta}$ and
$\boldsymbol{x}$.
\end{IEEEproof}
In the rest of the proof, we will focus on proving (\ref{eq:limtr}).
The proof for (\ref{eq:limtf}) and (\ref{eq:limtmu}) is similar.
We first show that for any positive integer $J>0$, we almost surely
have
\begin{align}
\lim_{t\rightarrow\infty}\left|\hat{r}_{k}^{t}-\overline{r}_{k}\left(\boldsymbol{\theta}^{t},\Omega^{J}\left(\boldsymbol{\mu}^{t},\boldsymbol{\theta}^{t}\right)\right)\right| & \leq\overline{e}_{J},\forall k,\label{eq:convftt}
\end{align}
where $\overline{e}_{J}$ satisfies $\lim_{J\rightarrow\infty}\overline{e}_{J}=0$. 

\textbf{Step 1 of proving (\ref{eq:convftt}):} Define a sequence
\begin{align}
\tilde{r}_{k}^{t} & =\left(1-\rho_{t}\right)\tilde{r}_{k}^{t-1}\nonumber \\
 & +\rho_{t}\sum_{i=tT_{s}+1}^{\left(t+1\right)T_{s}}\frac{r_{k}(\boldsymbol{\theta}^{t},\boldsymbol{x}^{J}(\boldsymbol{\mu}^{t},\boldsymbol{\theta}^{t},\boldsymbol{H}(i));\boldsymbol{H}(i))}{T_{s}}.\label{eq:fhd}
\end{align}
Comparing (\ref{eq:fit}) and (\ref{eq:fhd}), the update term $\sum_{i=tT_{s}+1}^{\left(t+1\right)T_{s}}\frac{r_{k}\left(\boldsymbol{\theta}^{t},\boldsymbol{x}^{J}\left(\boldsymbol{\mu}^{t},\boldsymbol{\theta}^{t},\boldsymbol{H}(i)\right);\boldsymbol{H}(i)\right)}{T_{s}}$
in (\ref{eq:fhd}) is only different from (\ref{eq:fit}) by $e_{J,t}=\left|\frac{\sum_{i=tT_{s}+1}^{\left(t+1\right)T_{s}}(r_{k}(\boldsymbol{\theta}^{t},\boldsymbol{x}^{J}\left(\boldsymbol{\mu}^{t},\boldsymbol{\theta}^{t},\boldsymbol{H}(i)\right);\boldsymbol{H}(i))-r_{k}\left(\boldsymbol{\theta}^{t},\boldsymbol{x}(i);\boldsymbol{H}(i)\right))}{T_{s}}\right|$.
Then we have
\begin{align}
\lim_{t\rightarrow\infty}\left|\tilde{r}_{k}^{t}-\overline{r}_{k}\left(\boldsymbol{\theta}^{t},\Omega^{J}\left(\boldsymbol{\mu}^{t},\boldsymbol{\theta}^{t}\right)\right)\right| & =0.\label{eq:convftt-1}
\end{align}
This is a consequence of \cite{Ruszczyski_MP80_SPthem}, Lemma 1,
which provides a general convergence result for any sequences of random
vectors $\left\{ \boldsymbol{\eta}^{t}\right\} ,\left\{ \boldsymbol{z}^{t}\right\} $
that satisfies conditions (a) to (e) in this lemma. When applying
\cite{Ruszczyski_MP80_SPthem}, Lemma 1 to prove the convergence of
$\tilde{r}_{k}^{t}$ in (\ref{eq:convftt-1}), we let $\boldsymbol{\eta}^{t}=\overline{r}_{k}\left(\boldsymbol{\theta}^{t},\Omega^{J}\left(\boldsymbol{\mu}^{t},\boldsymbol{\theta}^{t}\right)\right)$,
$\boldsymbol{z}^{t}=\tilde{r}_{k}^{t}$ and $\boldsymbol{\zeta}^{t}=\sum_{i=tT_{s}+1}^{\left(t+1\right)T_{s}}\frac{r_{k}(\boldsymbol{\theta}^{t},\boldsymbol{x}^{J}(\boldsymbol{\mu}^{t},\boldsymbol{\theta}^{t},\boldsymbol{H}(i));\boldsymbol{H}(i))}{T_{s}}$.
Since the instantaneous rate $r_{k}$ is bounded, we can find a convex
and closed box region $\mathcal{Z}$ to contain $r_{k}$ such that
condition (a) and (b) are satisfied. Since $\mathbb{E}\left[\boldsymbol{\zeta}^{t}\right]=\overline{r}_{k}\left(\boldsymbol{\theta}^{t},\Omega^{J}\left(\boldsymbol{\mu}^{t},\boldsymbol{\theta}^{t}\right)\right)$,
we have $\boldsymbol{b}^{t}=\boldsymbol{0}$ in condition (c) and
it follows from $\sum_{t=0}^{\infty}\left(\rho^{t}\right)^{2}<\infty$
that condition (c) is satisfied. Condition (d) follows from the assumption
on $\left\{ \rho^{t}\right\} $. Finally, from Lemma \ref{lem:Lipschitz-continuity-ofyf},
we have
\begin{align*}
 & \lim_{t\rightarrow\infty}\frac{|\overline{r}_{k}(\boldsymbol{\theta}^{t+1},\Omega^{J}\left(\boldsymbol{\mu}^{t+1},\boldsymbol{\theta}^{t+1}\right))-\overline{r}_{k}(\boldsymbol{\theta}^{t},\Omega^{J}\left(\boldsymbol{\mu}^{t},\boldsymbol{\theta}^{t}\right))|}{\rho_{t}}\\
\leq & \lim_{t\rightarrow\infty}O\left(\frac{\gamma^{t}B^{J}}{\rho_{t}}\right)=0,
\end{align*}
where the last inequality follows from (\ref{eq:rcont}) and $\sqrt{\left\Vert \boldsymbol{\mu}^{t+1}-\boldsymbol{\mu}^{t}\right\Vert ^{2}+\left\Vert \boldsymbol{\theta}^{t+1}-\boldsymbol{\theta}^{t}\right\Vert ^{2}}=O\left(\gamma_{t}\right)$.
Therefore, condition (e) in \cite{Ruszczyski_MP80_SPthem}, Lemma
1 is also satisfied, and thus (\ref{eq:convftt-1}) follows immediately
from \cite{Ruszczyski_MP80_SPthem}, Lemma 1.

\textbf{Step 2 of proving (\ref{eq:convftt}): }By the definitions
of $\tilde{r}_{k}^{t}$ and $\hat{r}_{k}^{t}$, we have
\begin{align*}
 & \left|\hat{r}_{k}^{t}-\tilde{r}_{k}^{t}\right|\\
\leq & \sum_{t^{'}=1}^{t}\left(1-\rho_{t}\right)^{t-t^{'}}\rho_{t^{'}}e_{J,t^{'}}\\
= & \sum_{t^{'}=1}^{n_{t}}\left(1-\rho_{t}\right)^{t-t^{'}}\rho_{t^{'}}e_{J,t^{'}}+\sum_{t^{'}=n_{t}+1}^{t}\left(1-\rho_{t}\right)^{t-t^{'}}\rho_{t^{'}}e_{J,t^{'}}\\
\leq & \rho_{1}e_{J,t}^{a}\frac{\left(1-\rho_{t}\right)^{t-n_{t}}}{\rho_{t}}+\frac{\rho_{n_{t}+1}}{\rho_{t}}e_{J,t}^{b},
\end{align*}
where $n_{t}=\left(1-\kappa-\epsilon\right)t$ with $\epsilon\in\left(0,1-\kappa\right)$,
$e_{J,t}^{a}=\max_{t^{'}\in\left\{ 1,...,n_{t}\right\} }e_{J,t^{'}}$
and $e_{J,t}^{b}=\max_{t^{'}\in\left\{ n_{t}+1,...,t\right\} }e_{J,t^{'}}$.
From Assumption \ref{asm:convS}-2, we have $\lim_{t\rightarrow\infty}\rho_{1}e_{J,t}^{a}\frac{\left(1-\rho_{t}\right)^{t-n_{t}}}{\rho_{t}}=0$
and $\lim_{t\rightarrow\infty}\frac{\rho_{n_{t}+1}}{\rho_{t}}<\infty$.
Moreover, from Theorem \ref{thm:Convergence-of-Short-term}, we have
$\lim_{J\rightarrow\infty}e_{J,t}^{b}=0,\forall t$. Then it follows
from the above analysis that $\lim_{t\rightarrow\infty}\left|\hat{r}_{k}^{t}-\tilde{r}_{k}^{t}\right|\leq\overline{e}_{J}$
for some error $\overline{e}_{J}$ satisfying $\lim_{J\rightarrow\infty}\overline{e}_{J}=0$.
Together with (\ref{eq:convftt-1}), it follows that (\ref{eq:convftt})
holds.

Then, we prove (\ref{eq:limtr}). From Theorem \ref{thm:Convergence-of-Short-term},
we have
\begin{equation}
\lim_{t\rightarrow\infty}\left|\overline{r}_{k}\left(\boldsymbol{\theta}^{t},\Omega^{J_{t}}\left(\boldsymbol{\mu}^{t},\boldsymbol{\theta}^{t}\right)\right)-\overline{r}_{k}\left(\boldsymbol{\theta}^{t},\Omega^{J}\left(\boldsymbol{\mu}^{t},\boldsymbol{\theta}^{t}\right)\right)\right|=\overline{e}_{J},\label{eq:dltr}
\end{equation}
for some error $\overline{e}_{J}$ satisfying $\lim_{J\rightarrow\infty}\overline{e}_{J}=0$.
Note that (\ref{eq:convftt}) holds for any finite $J$. Therefore,
for any $\epsilon>0$, there exists sufficiently large but finite
$J$ such that $\overline{e}_{J}\leq\frac{\epsilon}{2}$. Then it
follows from (\ref{eq:convftt}) and (\ref{eq:dltr}) that
\begin{align}
\lim_{t\rightarrow\infty}\left|\hat{r}_{k}^{t}-\overline{r}_{k}\left(\boldsymbol{\theta}^{t},\Omega^{J_{t}}\left(\boldsymbol{\mu}^{t},\boldsymbol{\theta}^{t}\right)\right)\right| & \leq\epsilon,\forall k.\label{eq:rfepson}
\end{align}
Since (\ref{eq:rfepson}) holds for any $\epsilon>0$, we have $\lim_{t\rightarrow\infty}\left|\hat{r}_{k}^{t}-\overline{r}_{k}\left(\boldsymbol{\theta}^{t},\Omega^{J_{t}}\left(\boldsymbol{\mu}^{t},\boldsymbol{\theta}^{t}\right)\right)\right|=0$. 

Finally, (\ref{eq:ghfhead}) follows directly from (\ref{eq:limtr})
- (\ref{eq:limtmu}) and the definition of $\hat{f}\left(\boldsymbol{\theta}\right)$.

\subsection{Proof of Theorem \ref{thm:Convergence-of-Algorithm1}\label{subsec:Proof-of-Theorem}}

Let $\boldsymbol{\phi}=\left[\boldsymbol{\theta}^{T},\boldsymbol{\mu}^{T}\right]^{T}$
denote the composite long-term control variables. For any $\overline{t}>0$,
we use $g\left(\boldsymbol{\phi}\right)$ as an abbreviation for $g\left(\overline{\boldsymbol{r}}\left(\boldsymbol{\theta},\Omega^{J_{\overline{t}}}\left(\boldsymbol{\mu},\boldsymbol{\theta}\right)\right)\right)$,
$\overline{\boldsymbol{r}}^{t}$ as an abbreviation for $\overline{\boldsymbol{r}}\left(\boldsymbol{\theta}^{t},\Omega^{J_{\overline{t}}}\left(\boldsymbol{\mu}^{t},\boldsymbol{\theta}^{t}\right)\right)$,
and $\overline{\boldsymbol{r}}^{t+1,t}$ as an abbreviation for $\overline{\boldsymbol{r}}\left(\boldsymbol{\theta}^{t+1},\Omega^{J_{\overline{t}}}\left(\boldsymbol{\mu}^{t},\boldsymbol{\theta}^{t}\right)\right)$,
when there is no ambiguity. 

1. We first prove that $\liminf_{t\rightarrow\infty}\left\Vert \bar{\boldsymbol{\phi}}^{t}-\boldsymbol{\phi}^{t}\right\Vert =0$
w.p.1. 

Since $\bar{f}^{t}\left(\boldsymbol{\theta}\right)$ is uniformly
strongly concave, we have
\begin{equation}
\nabla^{T}\bar{f}^{t}\left(\boldsymbol{\theta}^{t}\right)\boldsymbol{d}^{t}\geq\eta\left\Vert \boldsymbol{d}^{t}\right\Vert ^{2}+\bar{f}^{t}\left(\bar{\boldsymbol{\theta}}^{t}\right)-\bar{f}^{t}\left(\boldsymbol{\theta}^{t}\right)\geq\eta\left\Vert \boldsymbol{d}^{t}\right\Vert ^{2},\label{eq:ftdbound-1}
\end{equation}
for some $\eta>0$, where $\boldsymbol{d}^{t}=\bar{\boldsymbol{\theta}}^{t}-\boldsymbol{\theta}^{t}$.
Moreover, we have 
\begin{align}
g\left(\boldsymbol{\phi}^{t+1}\right) & \overset{\text{a}}{\geq}g\left(\overline{\boldsymbol{r}}^{t+1,t}\right)-o\left(\gamma^{t}\right)\nonumber \\
 & \overset{\text{b}}{\geq}g\left(\boldsymbol{\phi}^{t}\right)+\gamma^{t}\nabla_{\boldsymbol{\theta}}^{T}g\left(\boldsymbol{\phi}^{t}\right)\boldsymbol{d}^{t}\nonumber \\
 & -L_{0}\left(\gamma^{t}\right)^{2}\left\Vert \boldsymbol{d}^{t}\right\Vert ^{2}-o\left(\gamma^{t}\right)\nonumber \\
 & =g\left(\boldsymbol{\phi}^{t}\right)-L_{0}\left(\gamma^{t}\right)^{2}\left\Vert \boldsymbol{d}^{t}\right\Vert ^{2}-o\left(\gamma^{t}\right)\nonumber \\
 & +\gamma^{t}\left(\nabla_{\boldsymbol{\theta}}^{T}g\left(\boldsymbol{\phi}^{t}\right)-\nabla^{T}\bar{f}^{t}\left(\boldsymbol{\theta}^{t}\right)+\nabla^{T}\bar{f}^{t}\left(\boldsymbol{\theta}^{t}\right)\right)\boldsymbol{d}^{t}\nonumber \\
 & \geq g\left(\boldsymbol{\phi}^{t}\right)+\gamma^{t}\eta\left\Vert \boldsymbol{d}^{t}\right\Vert ^{2}-o\left(\gamma^{t}\right),\label{eq:keineq}
\end{align}
where $o\left(\gamma^{t}\right)/\gamma^{t}\rightarrow0$ as $t,\overline{t}\rightarrow\infty$
and $\nabla_{\boldsymbol{\theta}}g\left(\boldsymbol{\phi}^{t}\right)=\nabla_{\boldsymbol{\theta}}g\left(\overline{\boldsymbol{r}}\left(\boldsymbol{\theta}^{t},\Omega^{J_{\overline{t}}}\left(\boldsymbol{\mu}^{t},\boldsymbol{\theta}^{t}\right)\right)\right)$;
(\ref{eq:keineq}-a) follows from the first order Taylor expansion
$g\left(\overline{\boldsymbol{r}}^{t+1,t}\right)-g\left(\overline{\boldsymbol{r}}^{t+1}\right)=\nabla_{\overline{\boldsymbol{r}}}^{T}g\left(\overline{\boldsymbol{r}}^{t+1}\right)\left(\overline{\boldsymbol{r}}^{t+1,t}-\overline{\boldsymbol{r}}^{t+1}\right)+o\left(\gamma^{t}\right)$,
the fact that $\left\Vert \boldsymbol{\mu}^{t+1}-\nabla_{\overline{\boldsymbol{r}}}g\left(\overline{\boldsymbol{r}}^{t+1}\right)\right\Vert =\overline{e}_{\overline{t}}$
with $\lim_{\overline{t}\rightarrow\infty}\overline{e}_{\overline{t}}=0$,
and the definitions of $\overline{\boldsymbol{r}}^{t+1,t},\overline{\boldsymbol{r}}^{t+1}$;
(\ref{eq:keineq}-b) follows from the fact that the partial derivative
$\nabla_{\boldsymbol{\theta}}g\left(\boldsymbol{\phi}^{t}\right)$
is Lipschitz continuous with $L_{0}>0$ denoting the Lipschitz constant;
and the last inequality follows from (\ref{eq:ftdbound-1}) and $\lim_{t\rightarrow\infty}\left\Vert \nabla_{\boldsymbol{\theta}}g\left(\boldsymbol{\phi}^{t}\right)-\nabla\bar{f}^{t}\left(\boldsymbol{\theta}^{t}\right)\right\Vert =\overline{e}_{\overline{t}}$.
Let us show by contradiction that w.p.1. $\liminf_{t\rightarrow\infty}\left\Vert \bar{\boldsymbol{\phi}}^{t}-\boldsymbol{\phi}^{t}\right\Vert =0$.
Suppose $\liminf_{t\rightarrow\infty}\left\Vert \bar{\boldsymbol{\phi}}^{t}-\boldsymbol{\phi}^{t}\right\Vert \geq\chi>0$
with a positive probability. Then we can find a realization such that
$\left\Vert \boldsymbol{d}^{t}\right\Vert \geq\chi$ for all $t$.
We focus next on such a realization. By choosing a sufficiently large
$t_{0}$ and $\overline{t}$, there exists $\overline{\eta}>0$ such
that
\begin{align}
g\left(\boldsymbol{\phi}^{t+1}\right)-g\left(\boldsymbol{\phi}^{t}\right) & \geq\gamma^{t}\overline{\eta}\left\Vert \boldsymbol{d}^{t}\right\Vert ^{2},\forall t\geq t_{0}.\label{eq:gapf0xt}
\end{align}
It follows from (\ref{eq:gapf0xt}) that 
\[
g\left(\boldsymbol{\phi}^{t}\right)-g\left(\boldsymbol{\phi}^{t_{0}}\right)\geq\overline{\eta}\chi^{2}\sum_{j=t_{0}}^{t}\left(\gamma^{j}\right)^{2},
\]
which, in view of $\sum_{j=t_{0}}^{\infty}\left(\gamma^{j}\right)^{2}=\infty$,
contradicts the boundedness of $\left\{ g\left(\boldsymbol{\phi}^{t}\right)\right\} $.
Therefore, it must be $\liminf_{t\rightarrow\infty}\left\Vert \bar{\boldsymbol{\phi}}^{t}-\boldsymbol{\phi}^{t}\right\Vert =0$
w.p.1.

2. Then we prove that $\limsup_{t\rightarrow\infty}\left\Vert \bar{\boldsymbol{\phi}}^{t}-\boldsymbol{\phi}^{t}\right\Vert =0$
w.p.1. We first prove a useful lemma.
\begin{lem}
\label{lem:gapxbar}There exists a constant $\hat{L}>0$ such that
\[
\left\Vert \bar{\boldsymbol{\phi}}^{t_{1}}-\bar{\boldsymbol{\phi}}^{t_{2}}\right\Vert \leq\hat{L}\left\Vert \boldsymbol{\phi}^{t_{1}}-\boldsymbol{\phi}^{t_{2}}\right\Vert +e\left(t_{1},t_{2}\right),
\]
where $\lim_{t_{1},t_{2}\rightarrow\infty}e\left(t_{1},t_{2}\right)=0$.
\end{lem}

\begin{IEEEproof}
Following a similar analysis to that in Appendix \ref{subsec:Proof-of-Lemma},
it can be shown that
\begin{align}
\lim_{t\rightarrow\infty}\left|\bar{f}^{t}\left(\boldsymbol{\theta}\right)-\bar{g}_{\overline{t}}\left(\boldsymbol{\theta};\boldsymbol{\phi}^{t}\right)\right| & =O\left(\overline{e}_{\overline{t}}\right),\label{eq:fibconv}\\
\lim_{t\rightarrow\infty}\left\Vert \bar{\boldsymbol{\mu}}^{t}-\nabla_{\overline{\boldsymbol{r}}}g\left(\overline{\boldsymbol{r}}^{t}\right)\right\Vert  & =O\left(\overline{e}_{\overline{t}}\right),\label{eq:utappro}
\end{align}
where $\bar{g}_{\overline{t}}\left(\boldsymbol{\theta};\boldsymbol{\phi}^{t}\right)\triangleq g\left(\overline{\boldsymbol{r}}^{t}\right)+\nabla_{\boldsymbol{\theta}}g\left(\boldsymbol{\phi}^{t}\right)\left(\boldsymbol{\theta}-\boldsymbol{\theta}^{t}\right)-\tau\left\Vert \boldsymbol{\theta}-\boldsymbol{\theta}^{t}\right\Vert ^{2}$
and $\lim_{\overline{t}\rightarrow\infty}\overline{e}_{\overline{t}}\rightarrow0$.
It can be verified that $\bar{g}_{\overline{t}}\left(\boldsymbol{\theta};\boldsymbol{\phi}^{t}\right)$
and $\nabla_{\overline{\boldsymbol{r}}}g\left(\overline{\boldsymbol{r}}^{t}\right)$
are Lipschitz continuous in $\boldsymbol{\phi}^{t}$, and thus
\begin{align}
\left|\bar{g}_{\overline{t}}\left(\boldsymbol{\theta};\boldsymbol{\phi}^{t_{1}}\right)-\bar{g}_{\overline{t}}\left(\boldsymbol{\theta};\boldsymbol{\phi}^{t_{2}}\right)\right| & \leq B\left\Vert \boldsymbol{\phi}^{t_{1}}-\boldsymbol{\phi}^{t_{2}}\right\Vert ,\label{eq:LCfibar}\\
\left\Vert \nabla_{\overline{\boldsymbol{r}}}g\left(\overline{\boldsymbol{r}}^{t_{1}}\right)-\nabla_{\overline{\boldsymbol{r}}}g\left(\overline{\boldsymbol{r}}^{t_{2}}\right)\right\Vert  & \leq B\left\Vert \boldsymbol{\phi}^{t_{1}}-\boldsymbol{\phi}^{t_{2}}\right\Vert ,\forall\boldsymbol{\theta}\in\Theta\label{eq:mugbar}
\end{align}
for some constant $B>0$. Combining (\ref{eq:fibconv}) to (\ref{eq:mugbar}),
we have
\begin{align}
\left|\bar{f}^{t_{1}}\left(\boldsymbol{\theta}\right)-\bar{f}^{t_{2}}\left(\boldsymbol{\theta}\right)\right| & \leq B\left\Vert \boldsymbol{\theta}^{t_{1}}-\boldsymbol{\theta}^{t_{2}}\right\Vert +O\left(\overline{e}_{\overline{t}}\right)+e\left(t_{1},t_{2}\right),\label{eq:ft1t2}\\
\left\Vert \bar{\boldsymbol{\mu}}^{t_{1}}-\bar{\boldsymbol{\mu}}^{t_{2}}\right\Vert  & \leq B\left\Vert \boldsymbol{\theta}^{t_{1}}-\boldsymbol{\theta}^{t_{2}}\right\Vert +O\left(\overline{e}_{\overline{t}}\right)+e\left(t_{1},t_{2}\right),\label{eq:mubar}
\end{align}
where $\lim_{t_{1},t_{2}\rightarrow\infty}e\left(t_{1},t_{2}\right)=0$.
Since (\ref{eq:ft1t2}) holds for any $\overline{t}>0$ and $\lim_{\overline{t}\rightarrow\infty}\overline{e}_{\overline{t}}=0$,
we have
\begin{equation}
\left|\bar{f}^{t_{1}}\left(\boldsymbol{\theta}\right)-\bar{f}^{t_{2}}\left(\boldsymbol{\theta}\right)\right|\leq B\left\Vert \boldsymbol{\theta}^{t_{1}}-\boldsymbol{\theta}^{t_{2}}\right\Vert +e\left(t_{1},t_{2}\right),\forall\boldsymbol{\theta}\in\Theta.\label{eq:gapft1t2}
\end{equation}
Then it follows from (\ref{eq:gapft1t2}) and the Lipschitz continuity
and strong convexity of $\bar{f}^{t}\left(\boldsymbol{x}\right)$
that
\begin{equation}
\left\Vert \bar{\boldsymbol{\theta}}^{t_{1}}-\bar{\boldsymbol{\theta}}^{t_{2}}\right\Vert \leq B_{1}B\left\Vert \boldsymbol{\theta}^{t_{1}}-\boldsymbol{\theta}^{t_{2}}\right\Vert +B_{1}e\left(t_{1},t_{2}\right),\label{eq:xt1t2}
\end{equation}
for some constant $B_{1},B_{2}>0$. This is because for strictly convex
problem, when the objective function (\ref{eq:Pitert}) is changed
by amount $e\left(\boldsymbol{\theta}\right)$, the optimal solution
$\bar{\boldsymbol{\theta}}^{t}$ will be changed by the same order
(i.e., $\pm O\left(\left|e\left(\boldsymbol{\theta}\right)\right|\right)$).
Finally, Lemma \ref{lem:gapxbar} follows from (\ref{eq:mugbar})
and (\ref{eq:xt1t2}).
\end{IEEEproof}

Using Lemma \ref{lem:gapxbar} and following the same analysis as
that in \cite{Yang_TSP2016_SSCA}, Proof of Theorem 1, it can be shown
that \textbf{$\limsup_{t\rightarrow\infty}\left\Vert \bar{\boldsymbol{\phi}}^{t}-\boldsymbol{\phi}^{t}\right\Vert =0$
}w.p.1. Therefore, we have \textbf{
\begin{equation}
\lim_{t\rightarrow\infty}\left\Vert \bar{\boldsymbol{\phi}}^{t}-\boldsymbol{\phi}^{t}\right\Vert =0,\text{ w.p.1.}\label{eq:faigap}
\end{equation}
}

3. Finally, we are ready to prove the convergence theorem. By definition,
we have $\lim_{j\rightarrow\infty}\bar{\boldsymbol{\mu}}^{t_{j}}=\nabla_{\overline{\boldsymbol{r}}}g\left(\overline{\boldsymbol{r}}^{*}\right)$.
Then it follows from (\ref{eq:faigap}) that $\boldsymbol{\mu}^{*}=\lim_{j\rightarrow\infty}\boldsymbol{\mu}^{t_{j}}=\nabla_{\overline{\boldsymbol{r}}}g\left(\overline{\boldsymbol{r}}^{*}\right)$.
According to (\ref{eq:Pitert}), Lemma \ref{lem:Convergence-of-Surro}
and (\ref{eq:faigap}), $\boldsymbol{\theta}^{*}$ must be the optimal
solution of the following convex optimization problem w.p.1.:
\begin{align}
\underset{\boldsymbol{\theta}\in\Theta}{\text{max}}\: & \hat{f}\left(\boldsymbol{\theta}\right).\label{eq:Piterthead}
\end{align}
From the first-order optimality condition, we have
\begin{equation}
\nabla^{T}\hat{f}\left(\boldsymbol{\theta}^{*}\right)\left(\boldsymbol{\theta}-\boldsymbol{\theta}^{*}\right)\leq0,\forall\boldsymbol{\theta}\in\Theta.\label{eq:KKTPhead}
\end{equation}
It follows from Lemma \ref{lem:Convergence-of-Surro} and (\ref{eq:KKTPhead})
that $\boldsymbol{\theta}^{*}$ also satisfies (\ref{eq:ooutconv}).
Finally, (\ref{eq:inerconv}) follows from $\boldsymbol{\mu}^{*}=\nabla_{\overline{\boldsymbol{r}}}g\left(\overline{\boldsymbol{r}}^{*}\right)$
and Theorem \ref{thm:Convergence-of-Short-term}. This completes the
proof.


\end{document}